\documentclass[11pt, oneside]{article}

\usepackage{type1cm}        %
\usepackage{makeidx}         %
\usepackage{graphicx}        %
\usepackage{multicol}        %
\usepackage[bottom]{footmisc}%

\usepackage[utf8]{inputenc} %

\usepackage{newtxtext}       %
\usepackage{newtxmath}       %

\usepackage{authblk}

\usepackage{siunitx}
\usepackage{notoccite}

\usepackage{float} %
\usepackage{wrapfig} %

\makeindex             %
\usepackage{todonotes}

\usepackage{authblk}

\usepackage{hyperref}

\DeclareRobustCommand{\orderof}{\ensuremath{\mathcal{O}}}

\title{\vspace{-4.0cm}The Tracking Machine Learning challenge : Accuracy phase\footnote{To be published in NeurIPS 2018 Competition Book, Springer Series on Challenges in Machine Learning}}

\author[1]{Sabrina Amrouche}
\author[2]{Laurent Basara}
\author[3]{Paolo Calafiura}
\author[2]{Victor Estrade}
\author[3]{Steven Farrell}
\author[4]{Diogo R. Ferreira}
\author[5]{Liam Finnie}
\author[6]{Nicole Finnie}
\author[2]{C\'ecile Germain}
\author[7]{Vladimir Vava Gligorov}
\author[1]{Tobias Golling}
\author[8]{Sergey Gorbunov}
\author[3]{Heather Gray}
\author[9]{Isabelle Guyon}
\author[10]{Mikhail Hushchyn}
\author[11]{Vincenzo Innocente}
\author[1]{Moritz Kiehn}
\author[12]{Edward Moyse}
\author[13]{Jean-Fran\c{c}ois Puget}
\author[14]{Yuval Reina}
\author[15]{David Rousseau \thanks{contact: rousseau@lal.in2p3.fr. D.R. Ferreira, S. Gorbunov, L. Finnie, N. Finnie, J.F. Puget, Y. Reina, J.S. Wind and T. Xylouris are participants, all others are organizers}}
\author[11]{Andreas Salzburger}
\author[10]{Andrey Ustyuzhanin}
\author[16]{Jean-Roch Vlimant}
\author[17]{Johan Sokrates Wind}
\author[18]{Trian Xylouris}
\author[15]{Yetkin Yilmaz}

\affil[1]{D\'epartement de Physique Nucl\'eaire et Corpusculaire, Universit\'e de Gen\`eve, Geneva, Switzerland}
\affil[2]{LRI/TAU, Univ. Paris-Sud/INRIA/CNRS, Universit\'{e} Paris-Saclay, Gif-sur-Yvette, France}
\affil[3]{Physics Division, Lawrence Berkeley National Laboratory and University of California, Berkeley CA, USA }
\affil[4]{IST, University of Lisbon, Lisbon, Portugal}
\affil[5]{IBM Germany Research and Development, Germany}
\affil[6]{Bosch Center for Artificial Intelligence, Germany}
\affil[7]{LPNHE, Sorbonne Universit{\'e}, Paris Diderot Sorbonne Paris Cit{\'e}, CNRS/IN2P3, Paris, France}
\affil[8]{Goethe University Frankfurt am Main, Germany}
\affil[9]{UPSud/INRIA Universit\'e Paris-Saclay, Orsay, France, and ChaLearn, Berkeley, CA, USA}
\affil[10]{National Research University Higher School of Economics and Yandex School of Data Analysis, Moscow, Russia} 
\affil[11]{CERN, Geneva, Switzerland} 
\affil[12]{Department of Physics, University of Massachusetts, Amherst MA, USA }
\affil[13]{Data and AI R\&D, IBM France Lab, Biot, France}
\affil[14]{Tel-Aviv, Israel}
\affil[15]{LAL, Univ. Paris-Sud, CNRS/IN2P3, Universit\'e Paris-Saclay, Orsay, France}
\affil[16]{California Institute of Technology, Pasadena CA, USA}
\affil[17]{Norwegian University of Science and Technology, Oslo, Norway }
\affil[18]{Frankfurt am Main, Germany }

\date{}

\begin{document}

\hypersetup{pageanchor=false}

\maketitle
\hypersetup{pageanchor=true}

\abstract{
This paper reports the results of an experiment in high energy
physics: using the power of the ``crowd" to solve difficult
experimental problems linked to tracking accurately the trajectory of particles in the Large Hadron Collider (LHC). This experiment took the form of a machine learning
challenge organized in 2018: the Tracking Machine Learning
Challenge (TrackML). Its results were discussed at the competition session at the Neural Information Processing Systems conference (NeurIPS 2018). Given 100.000 points, the participants had to connect them into about 10.000 arcs of circles, following the trajectory of particles issued from very high energy proton collisions.  The competition was difficult with a dozen front-runners well ahead of a pack. The single competition score is shown to be accurate and effective in selecting the best algorithms from the domain point of view. The competition has exposed a diversity of approaches, with various roles for Machine Learning, a number of which are discussed in the document. }

\section{Introduction}
\label{s:intro}

TrackML\cite{trackmlwebsite} is the third ML challenge for particle physics. After the success of the Higgs Boson challenge~\cite{adam-bourdarios_higgs_2014} and the Flavour of Physics challenge~\cite{flavourKaggle}, the goal was to address a completely different issue, which is critical to ensure the quality of novel particle detection at the Large Hadron Collider (LHC)\cite{CERNLHC} at CERN: Tracking accurately the trajectory of particles in the LHC detectors. The challenge was organized
by an interdisciplinary team of physicists from three LHC experiments (ATLAS, CMS, and LHCb), computer
scientists, and ChaLearn, a non-profit group dedicated to the organization
of challenges in Machine Learning, and supported by a number of sponsors listed in the Acknowledgments section.
No knowledge of particle physics was necessary to participate.

\begin{figure*}
\begin{center}
\includegraphics [width=\linewidth]{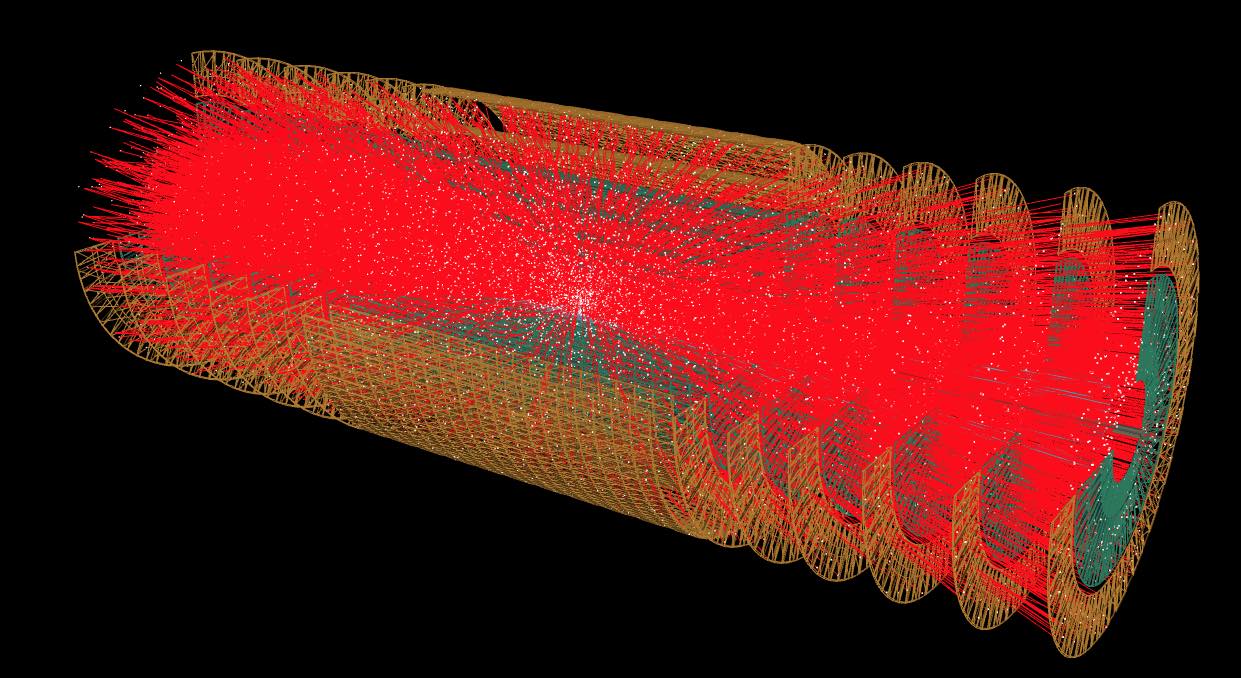}

\caption{TrackML detector (one sector of the detector has been etched out). White dots are the measured points, while the red lines are the trajectories of the particles.}
\label{fig:3d_display}
\end{center}
\end{figure*}

\begin{figure*}
\begin{center}
\includegraphics [width=5 cm]{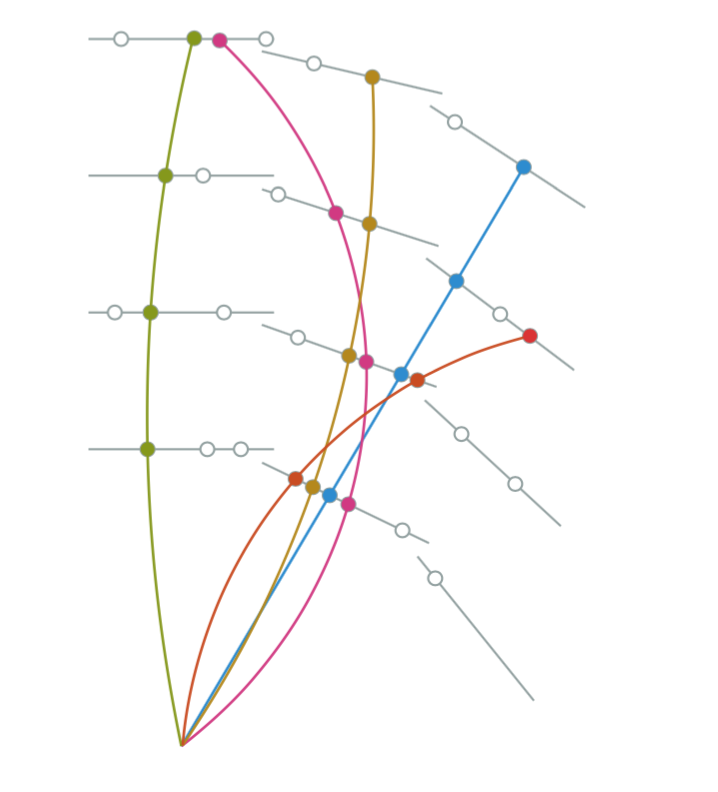}

\caption{A very simplified view in 2-Dimension : the name of the game is to associate the points into tracks.}
\label{fig:proj2D}
\end{center}
\end{figure*}

The LHC 
is a unique particle accelerator complex colliding protons at unprecedented energies. It allowed the Higgs boson discovery in 2012 (acknowledged by the 2013 Nobel prize in physics).
It will collect data of increasing complexity and at increasing rate with a large upgrade so called High Luminosity LHC planned for 2025.
All the analysis pipelines of the proton collisions (or {\it events}) rely on a first step, the reconstruction of the 3D trajectories of the particles within the detector (see Figure~\ref{fig:3d_display}) and Figure~\ref{fig:proj2D}. This problem is currently conveniently solved by combinatorial optimization methods (based on Kalman filters).
But the CPU time to reconstruct the trajectories (helices) from the measurements (3D points) is expected to increase faster than the projected computing resources. New approaches to pattern recognition are needed to exploit fully the discovery potential of the HL-LHC.

The overall goal of the challenge was to explore new methods to address the trade-off between algorithmic quality (good track reconstruction) and speed.
From the Machine Learning point of view, the problem can be treated as a {\bf latent variable problem} similar to clustering, in which particle trajectory ``memberships'' must be inferred, although the ratio between the number of clusters (10K) and their size (10 points), is highly unusual. 
It can also be treated as a {\bf tracking problem} considering trajectories as time series, or a {\bf pattern de-noising problem} considering that the dotted trajectories are noisy versions of continuous traces. One important point is that the points on one trajectory are not geometrically close to each other (a human cannot associate the points by eye), but they follow a specific pattern : a distorted arc of helix pointing approximately to the origin.

The HEP (High Energy Physics) experiments have embraced Machine Learning, originally for supervised classification as a tool in the final analysis stage, and for exploring more diverse applications. Recent attempts of applying Machine Learning to particle physics pattern recognition-tracking indicate a strong potential~\cite{heptrkx}. 
A one day hackathon\cite{TrackMLRamp2017}  limited to a two-dimensional problem has shown the richness of approaches and the setup to be tested.

A dataset consisting of an accurate simulation\cite{ACTS} of a LHC experiment has been created, listing for each event the measured 3D points, and the list of 3D points associated to a true track. The data set is large to allow the training of data hungry Machine Learning methods; the orders of magnitude are: ten thousand events, one billion points, on	e hundred million tracks. The participants to the challenge should find the tracks in an additional test dataset, meaning building the list of 3D points belonging to each track (deriving the track parameters is not the topic of the challenge). The emphasis is to expose innovative approaches, rather than hyper-optimizing known approaches. The challenge has been run in two phases:

\begin{enumerate}
\item During the first "Accuracy" phase, which has run on Kaggle \cite{TrackMLKaggle} from 1\textsuperscript{st} May 2018 to 10\textsuperscript{th} August 2018, a metric reflecting the accuracy of the model at finding the proper point association that matters to most physics analysis to identify the best algorithms. The metric is based on the overall fraction of points associated to a true track.
\item The second "Throughput" phase has been running from October 2019 to 12th March 2019 on Codalab~\cite{TrackMLCodalab}. It focuses on optimizing the inference speed, starting from the collection of algorithms exposed in the first phase. The training speed remains unconstrained. 
\end{enumerate}

As the second phase is still running at the time of writing, this document focuses on the first phase.

The success of the Challenge can be attributed in part to the 
visibility of CERN, and the appeal of the problem.  We also
stimulated participation by providing a starting kit, responding
promptly to questions in the online forum, where participants were
also helping each other, and through wide advertising. An additional
incentive was provided in the form of prizes for the winners and
an invitation to visit CERN to discuss their results with high-energy
physicists. Design choices also played an important role. Simplifying the problem setting in order to reach computer scientists, while keeping it realistic enough for the challenge to be useful, was much more difficult than for the HiggsML. We were largely successful, as the many solutions, including the winning ones, come from computer scientists. However, for most solutions, the focus has been mostly on optimization, rather than the variety of ML methods we envisioned.

Given the interest raised by the Challenge and the willingness to pursue the study beyond the formal end of the competition, 
a very similar dataset will be made permanently available on the CERN open data portal\cite{TrackMLCERNDataPortal} and on the UCI repository\cite{TrackMLUCI},
together with accompanying software and documentation.

The document is structured as follows: section~\ref{s:setup} details the setup of the competition, with in particular the dataset (\ref{s:dataset} and  the scoring algorithm \ref{s:scoring}).  Section~\ref{s:competition} summarizes the competition proceedings and studies devoted to the accuracy of the ranking. Section ~\ref{s:perf} presents the performances of selected algorithms while the long section~\ref{s:algos} gives a brief summary of the different techniques that have been used, before the conclusion and outlook in section~\ref{s:conclusion}.

\section{Setup}
\label{s:setup}

An event is a set of particle measurements in the detector. 
From an abstract point of view, the detector is simply an apparatus that records the impacts, called the hits, of the particles traversing the detector in an event, i.e. each time a pair of protons collides. The detector is formed of discrete layers. One event has $\orderof \left(10^5\right)$ hits, corresponding to $\orderof \left(10^4\right )$ particles.

The basic configuration is as follows.
\begin{itemize}
\item Each particle is created close to, but not exactly, at the center of the detector (see section~\ref{s:dataset} for details). 
\item  Each hit is a 3D measurement in Cartesian coordinates $(x,y,z)$. 
For each particle, the number of hits is on average 12, but as low as 4 and as large as 20. 
\item The participants should associate the hits created by each particle together, to form tracks. Typically, at least 90\%  of the true tracks should be recovered,
\item The tracks are slightly distorted arc of helices with axes parallel to the $z$-axis, and pointing approximately to the interaction center. 

\end{itemize}

In an ideal world:
\begin{itemize}
\item each particle would leave one and only one hit on each layer of the detector
\item the trajectories would be exact arcs of helices;
\item the  $(x,y,z)$ coordinates would be exact. 
\end{itemize}

In this ideal world, fitting the parameters of the helices suffices to solve the problem. However, there are a number of subtleties: 
\begin{itemize}
\item Depending of the local geometry, each particle may leave multiple hits in a layer, and the layer may not record anything at all. 
\item The arcs are often slightly distorted.  
\item The measurements have some non isotropic uncertainty.
\end{itemize}

The challenge is to show robustness of the algorithm with respect to all these perturbations. It is enforced by the metric defined by the score.

\subsection{The TrackML detector}
\label{s:detector}

The challenge relies on a realistic detector model to simulate
measured particle hits similar to what is expected for an HL-LHC
experiment. The detector model is inspired by the  ATLAS and CMS upgrade tracker designs
\cite{ATLAS-TDR-030,CMS-TDR-014}.
They are based on large surface all-silicon
detectors with a cylinder-like geometry in the central regions and a disk-like
geometry in the forward regions.

The coordinate system is a right-handed Cartesian coordinate system $(x,y,z)$ with the global $z$ axis defined along the beam direction, which is the axis of symmetry of the cylinders and disks composing the detector.
The $(x-y)$ plane is referred to as the transverse plane, and the azimuthal angle $\phi \in [-\pi,\pi[$ is defined in the transverse plane with $\phi=0$ denoting the $x$-axis. The polar angle $\theta$ is measured from the $z$-axis and is defined to be within $[0,\pi]$. 

\begin{figure}[htb]
\begin{center}
\includegraphics[width=0.9\linewidth]{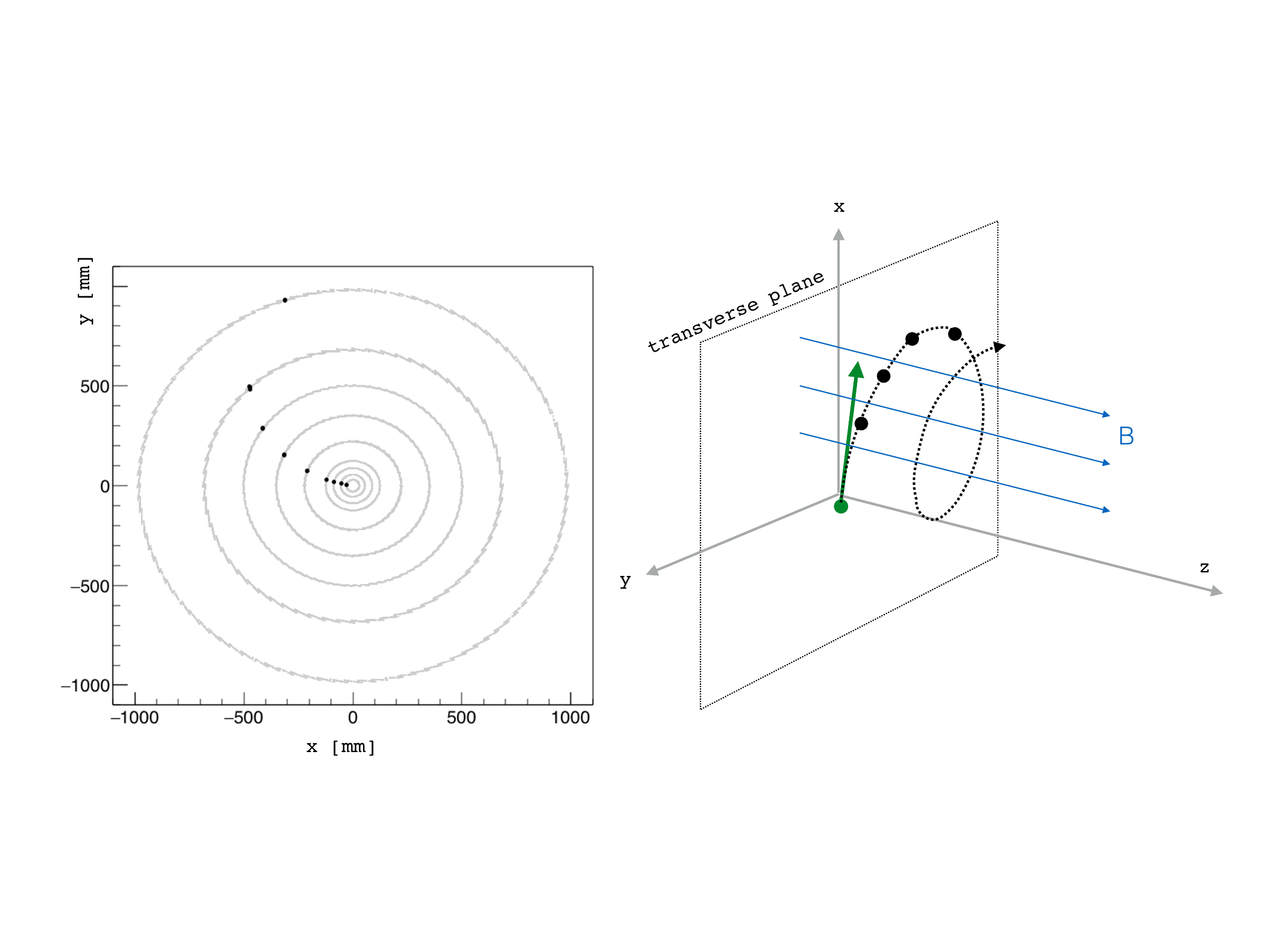}
\caption{Illustration of a particle moving through a constant magnetic field. In the transverse projection, this corresponds to the particle moving along a circle.}
\label{fig:helix}
\end{center}
\end{figure}

In order to measure the particle momentum, tracking detectors are embedded in a strong magnetic field.
A charged particle, when moving through a constant magnetic field follows a
helical trajectory (figures ~\ref{fig:helix}). 
The magnetic field is usually aligned with the beam direction, such that the particle is bent in the transverse plane.  

\begin{figure}[htb]
  \centering
  \resizebox{\linewidth}{!}{%
    \includegraphics[height=6cm]{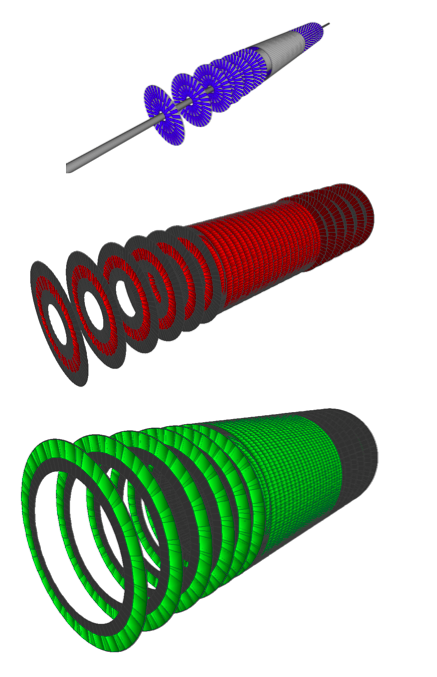}%
    \hspace{1cm}%
   \includegraphics[height=6cm]{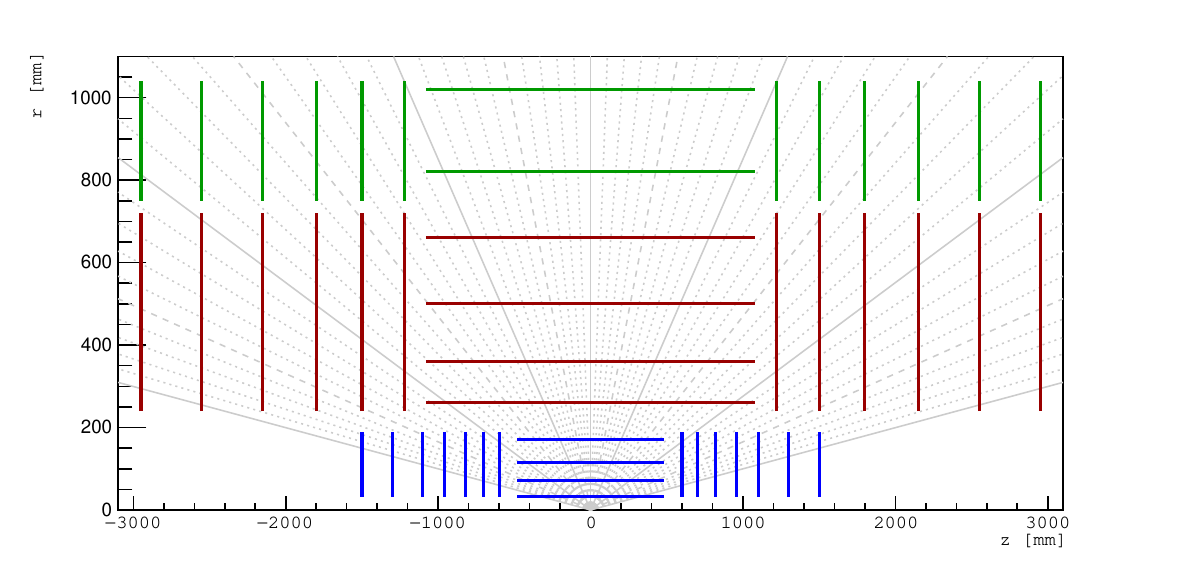}%
  }
  \caption{Detector layout for the virtual TrackML detector. On the left
    the three major sub-detectors, pixel, short strips, and long strips,
    are shown separately. On the right, a schematic of the full layout
    and its coverage along the radial and longitudinal dimensions as
    well as in the $\eta$ direction is shown. The different colors
    represent the different sub-detectors while the marked numbers are
    the internal volume and layer identifiers.}
  \label{fig:detector}
\end{figure}

The full detector geometry is shown in figure \ref{fig:detector}. The
detector is split into three separate sub-detectors that differ in
spatial resolution and passive material. The inner-most sub-detector is a
pixel detector with a spatial resolution of \SI{50x50}{\um} and further
out two different strip detectors with short \SI{80x1200}{\um} and long
strips \SI{0.12x10.8}{\mm} are placed. Each detector includes
realistic module geometries with placement and overlap chosen to yield a
hermetic coverage up to $\eta = \num{3}$.

\subsection{Simulation}
\label{s:simulation}

This section gives more details for reference on the physics involved in the simulation, non physicists can skip it.

The particle content of the collisions is generated using the Pythia 8 event generator \cite{SjostrandEtAl:2008:bip}. A hard Quantum Chromodynamics (QCD) interaction
that generates a $t\bar{t}$-pair is used as the signal. An additional
\num{200} soft QCD interactions are overlaid to simulate the expected
pile-up conditions at the HL-LHC. The interaction vertices are spread
out over a luminous region with a width of \SI{5.5}{\mm} along the beam
axis.

Charged particles are propagated through the detector using a fast
detector simulation based on the ACTS software
\cite{ACTS}. An inhomogeneous magnetic field
similar to the one in the ATLAS experiment is used and material
interactions, i.e. multiple scattering, energy loss, or hadronic
interactions, are simulated using parametric models. Only tracks with a
transverse momentum above \SI{150}{\MeV} are propagated. Inefficient
sensors and additional hits from noise or particles below the threshold
are also included. Most particles are primary particles produced near the origin. Primary particles can produce (through decay or interaction with the detector material itself) secondary particles which are hence created at a distance from the origin. 

\begin{figure}[ht]
\includegraphics[width=\textwidth]{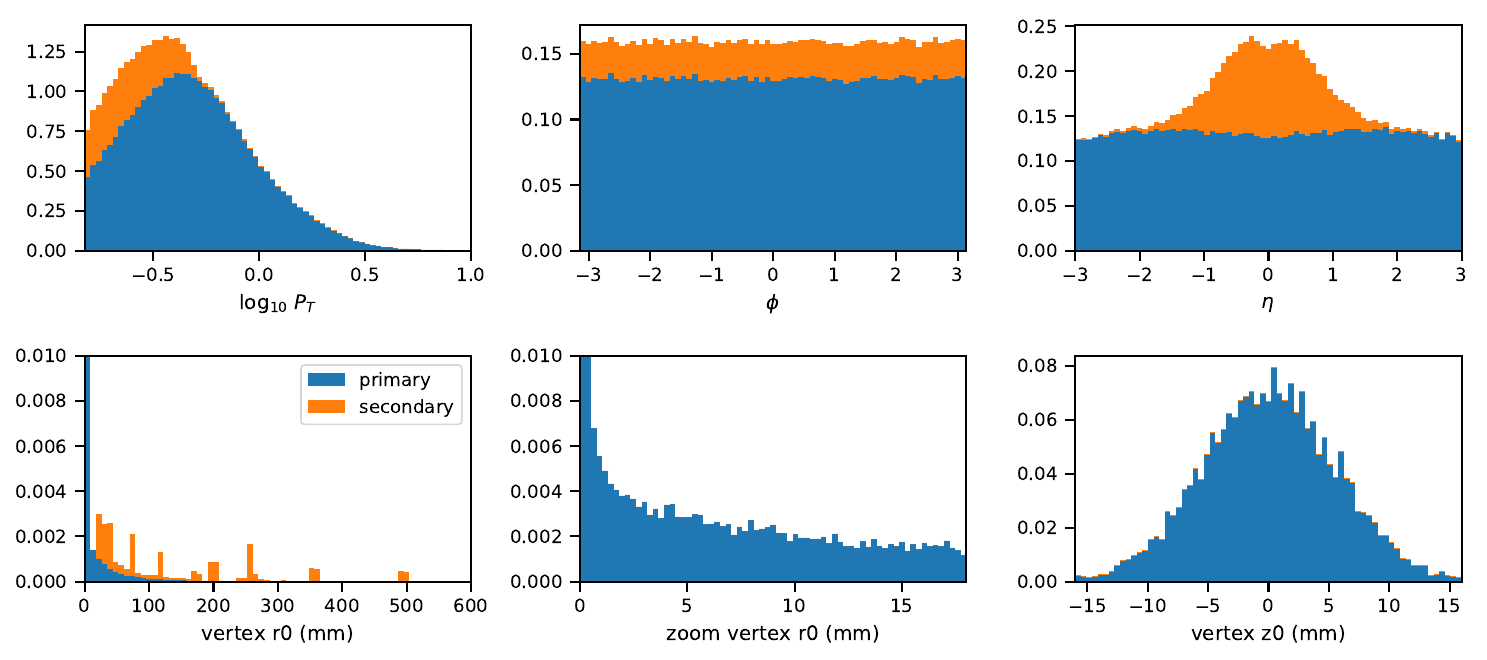}
\caption{Probability density function of primary (blue) and secondary (orange) particles, normalized to the sum of the distributions, as a function of six kinematic variables, in the coordinate system described in section~\ref{s:detector}.
$log_{10}(p_T)$ (top left)is the transverse momentum which is proportional to the radius of curvature, so that particles on the right side are almost straight.  
$\phi$ (top middle) is the azimuth angle.$\eta=\log(\tan(\theta/2))$ (top right) is the pseudo-rapidity, related to the dip angle $\theta$: $\eta$=0 for particle perpendicular to the z axis, +3 for particle very close to the z axis on the positive $z$ side, -3 on the negative side.
$r_0$ (bottom  left with a zoom bottom middle) is the transverse radius of the point of origin of the particle , and $z_0$ (bottom right) is the $z$ coordinate of the origin of the particle.}
\label{fig:unweightedpdf}       
\end{figure}

Figure~\ref{fig:unweightedpdf} shows various kinematic distributions of the particles.  There is no dependence on the $\phi$ angle, and the distribution of the primary particles as a function of $\eta$ is flat (it's a known properties of primary particles for this variable), unlike the secondary particles which peak around 0. The secondary particles have a lower momentum than the primaries and, being produced in the interaction of the primary particles with the matter of the detector, the transverse radius of the production vertex $r_0=\sqrt{x_0^2+y_0^2}$ corresponds to the location of the modules.

\subsection{Dataset}
\label{s:dataset}

Particle physics events contain a multitude of different types of
information that are usually represented in a nested structures of
variable length. Here, we aim to provide the data in a flattened
structure to avoid the necessity for specialized tools or formats.
Since events are statistically independent, data is stored separately
for each event, organized by a numerical event identifier.
For each event in the training dataset the following four files are provided in CSV format (only a brief description is given here, more details are available in the participant document\cite{TrackMLKaggle} and in the open data set description\cite{TrackMLCERNDataPortal}.
\begin{itemize}
    \item {\bf Hits} : Each entry has a unique identifier and provides the simulated hit information
  that are the core input for the challenge, essentially the coordinates $(x,y,z)$ 
    \item {\bf Cells} :  Each entry  enables
  participants to extract additional information, e.g. directional
  information 
    \item{\bf Hit truth} : Each entry has the same unique identifier as the hits file, gives the true hit position and which particle created the hit
    \item{\bf Particles truth} : Each entry has a unique id and  represents one generated, charged, final  state particle.
\end{itemize}

Training and test data-sets only differ by the type of files made
available. In the test data set only the hits and cells files are
available, as would be the case in the real data from the detector. Internally, all events were generated in exactly the
same way.

Participants had to provide their solutions also in CSV
format. Due to technical requirements by the Kaggle platform a single
file had to be provided for all events in the test dataset. Each entry
therefore had to comprise three values: the event identifier as
provided, the event-unique hit identifier as provided, and an arbitrary track
identifier generated by the participant. Hits reconstructed to belong to
the same track must have the same track identifier.
All the hits of all events should be listed once and only once.

\subsection{Scoring}\label{s:scoring}

When developing a tracking algorithm for particle physics, experts usually assemble a large number of histograms to assess its quality. However, as usual for a Kaggle competition, algorithms had to be ranked based on a single score to evaluate their quality. 
Participants submit a submission file, which proposes a partition of the hits into a list of suggested tracks. The score evaluates the quality of this partition by comparing to the ground truth partition.

In one sentence: the score is based on the intersection between the reconstructed tracks and the ground truth particles, normalized to one for each event, and averaged on the events of the test set. It is implemented in the helper library \cite{TrackMLlibrary}. More details follow.

Only particles  with four hits or more are considered, and only proposed tracks with four hits or more are considered. 
Each track is matched with the ground truth { \it majority particle} sharing with it the greatest hit number.
The ratio of this intersection to the number of hits of the reconstructed track defines the {\it track purity}, while the ratio of this intersection to the number of hits of the underlying particle defines {\it particle purity}. Both ratios have to be above 50\% to define a good track so that a one-to-one relationship between particle and track can be defined.

Each hit has a weight $w_i$, as explained below,  and the total score $S$ for a given solution
is given by the following multiple sum:

\begin{equation}
      S =  \frac{1}{N_{\text{events}}} \sum_{\{\text{events}\}} \sum_{\{\text{ good tracks}\}}
        \sum_{\{\text{intersection hits}\}} w_i 
\end{equation}

This score is largely consistent with the various and more complicated metrics used in physics reconstruction. It is a combination of the Jaccard version of counting pairs \cite{BenHur} and set matching~\cite{meila}. With set matching, it shares the one-to-one assignment of reconstructed clusters to true clusters. However, thanks to the majority rule, it does not suffer from the ``problem of matching''~\cite{meila}. With respect to counting pairs, the Jaccard index is more appropriate than the Rand index~\cite{Rand71}, as the result of the later would be dominated by the true negatives (pair of points that agree to be in different clusters), which are not taken into account in the Jaccard counting points index.

With this definition no penalty for incorrect hits is necessary (as usually done by physicists), since a wrongly associated hit will automatically reduce the score for the other track it should have been associated to.

As commonly done in Kaggle competitions, the participants' submissions are based on a test dataset of (here) 125~events. 
There is a secret split of these events into 36~events to be used to compute the score for the public leader board (updated online) and 89 to be used for the final private leader board which determined the final ranking at the end of the competition.

\begin{figure}[htb]
  \centering
  \resizebox{\linewidth}{!}{%
    \includegraphics[height=5cm]{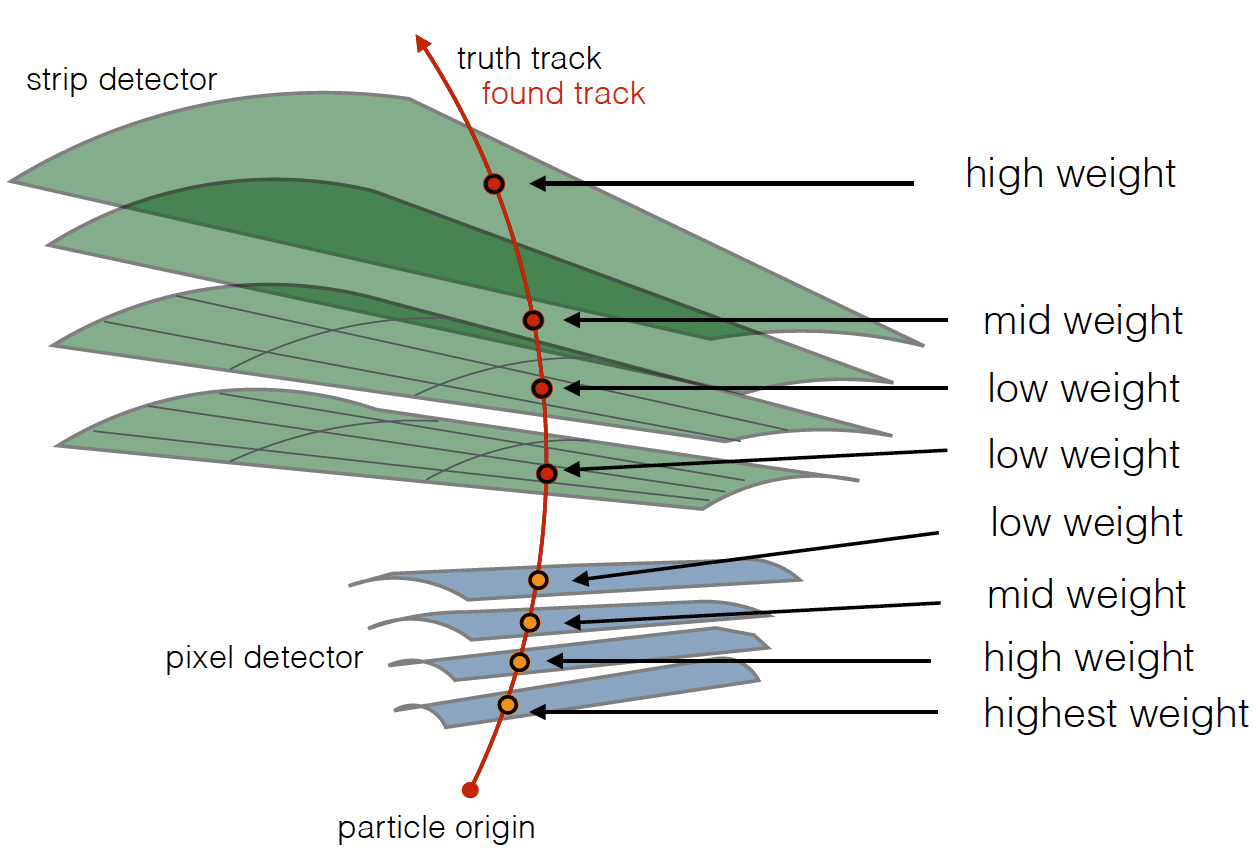}%
    \hspace{1cm}%
    \includegraphics[height=5cm]{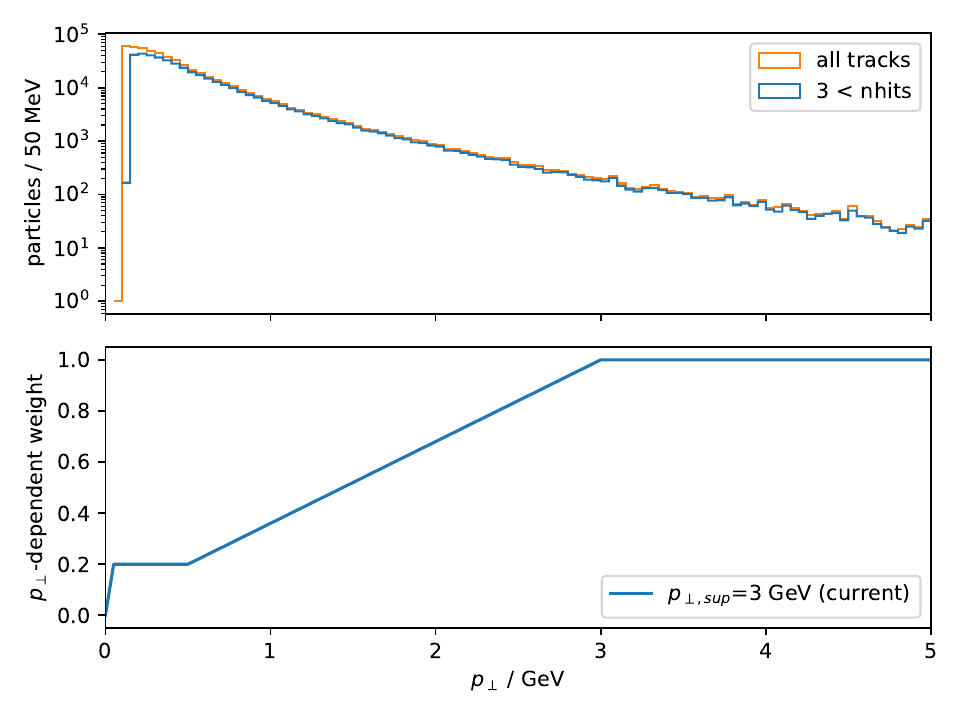}%
  }
  \caption{On the left: illustration of the order-dependent hit weight.
  On the right: the simulated $p_\perp$ distribution and the
  $p_\perp$-dependent hit weight.}
  \label{fig:hitweights}
\end{figure}

The contributions to the per-hit weight are illustrated in figure \ref{fig:hitweights}.
The order-dependent weight penalizes missed hits at the inner and outer-most part of the detector more strongly.
Missing a hit on the innermost layer will strongly influence the quality of the extrapolation of the track to the origin, while a missed hit on the outer layers reduces the lever arm for the momentum measurement.
The $p_\perp$-dependent part favors reconstructing high-momentum tracks over low-momentum tracks without excluding either region completely: high-momentum (straight) tracks are more likely to come from an interesting physics process but are more rare. Hence

The overall score is normalized such that a perfect algorithm has a score of one.
A random algorithm has a null score. A given algorithm may leave undecided a number of hits ; as a valid solution requires all hits to be assigned, these hits are usually assigned to a single "garbage" track with possibly thousands of hits ; this garbage track has a null contribution to the score. 

\begin{figure}[ht]
\includegraphics[width=\textwidth]{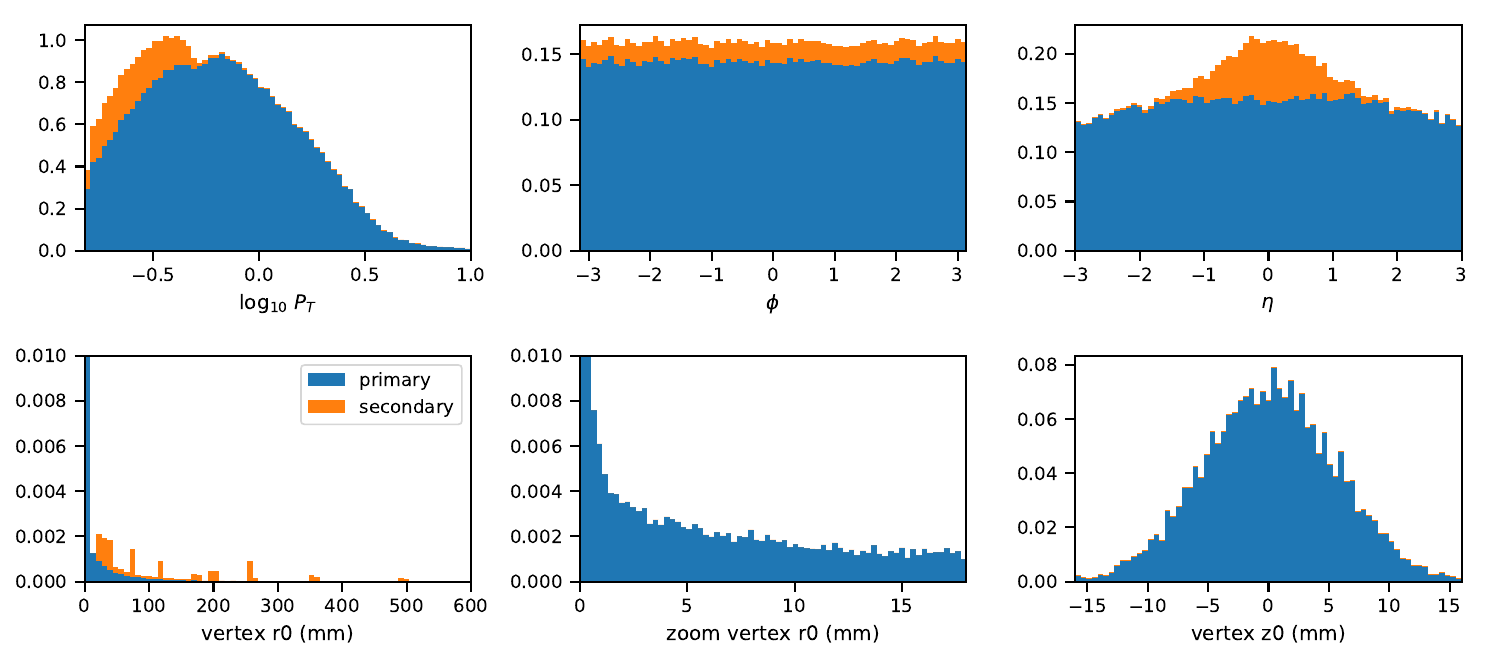}
\caption{Probability density functions taking into account the score weight. The variables are described in Fig.~\ref{fig:unweightedpdf}.
}
\label{fig:weightedpdf}       
\end{figure}

Figure~\ref{fig:weightedpdf} shows the weighted distribution of the particles, similar to Figure~\ref{fig:unweightedpdf} (which is not weighted). One sees that the weight has reduced the low momentum part and reduced the contribution of secondary particles.

\subsection{The Kaggle platform}

The accuracy phase uses the Kaggle platform
\cite{TrackMLKaggle}. This platform hosts the dataset and
provides the scoring and the leader-board for the
participants. Participants can download the training and test dataset,
train on the former and prepare a solution for the latter, and upload
the solution.

While a variety of solution metrics already existed, the scoring metric discussed in section \ref{s:scoring} was not one of them.
It was implemented by Kaggle on their platform specifically for this challenge.
For the test dataset, 125 simulated events were selected. Participants had to reconstruct all events on their own machines and upload their solution to Kaggle where the score was then computed.

\subsection{Summary of challenge choices}

When designing a challenge, the domain problem has to be simplified so that it can be tackled by participants of diverse backgrounds, often without any domain knowledge. The various simplification choices are listing below:
\begin{itemize}
\item the data comes from a simplified simulation with Acts, instead of real data from the detector or data from a detailed simulation \cite{ATLASsimulation} : the simplified simulation still has a number of relevant features which make track finding as difficult as in the detailed simulation or real data
\item simple geometry with modules arranged in cylinders and disks, instead of a more complex geometry with cones, and accurate simulation of electronics, cooling tubes and cables : there is no reason to believe the more complex geometry would lead to radically different algorithms
\item no merging of reconstructed hits : merging of hits would occur in less than 0.5\% of the cases. Dealing with such cases is eventually necessary but it would not radically change the core algorithm use, so the added complexity was deemed unnecessary
\item only one type of event (top quark simulation) instead of a variety of events : top quark simulation is often used for algorithm validation because of many possible final state including electron muon and $\tau$ lepton, b c and light quark
\item one single metric based on the hit clustering, while algorithms are evaluated with many metrics, in particular some based on the quality of the evaluation of the particle parameter from the associated hits : deriving the particle parameters from the associated hits is a well defined process done with proven techniques like Kalman filtering, so this aspect has been removed from the scope of the challenge. In addition, experience of the expert of the domain is that the quality of the particle parameter evaluation is directly impacted by the quality of the hit clustering. 
\end{itemize}

\subsection{Starting kit}
\label{s:startingkit}

The baseline solutions for the challenge were presented in the form of executable Jupyter notebooks~\cite{PER-GRA:2007}. Participants were introduced into three basic approaches that helped to develop geometrical intuition behind the task. 
The first approach was straightforward k-Nearest Neighbor classifier that was trained on a bunch of tracks and was able to discriminate hits that fall into the proximity of certain tracks venues. However, the performance of such an approach was quite limited. 

The second approach introduces data preprocessing based on the observation that most of the track hits follow an arc of a helix pattern: 

$$ 
r_{1} = \sqrt{x^{2}+y^{2}+z^{2}}
$$

\begin{figure}[ht]
\includegraphics[width=0.8\textwidth]{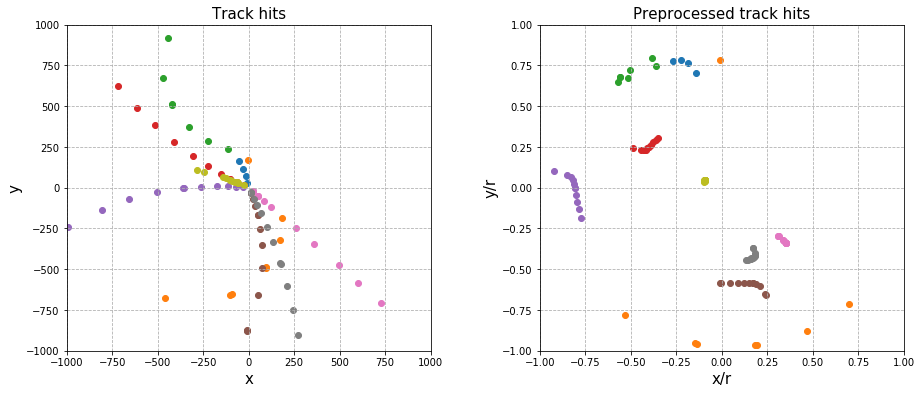}
\caption{Hit coordinates transformation before DBSCAN: original (left) and transformed (right).}
\label{fig:sk_hit_transform}       
\end{figure}

Hence we can transform hit coordinates in the following way (see Figure~\ref{fig:sk_hit_transform}) so the hits from the same helix would be close to each other after the mapping.

$$
x_{2} = x / r_{1}
$$

$$
y_{2} = y / r_{1}
$$

$$
r_{2} = \sqrt{x^{2}+y^{2}}
$$

$$
z_{2} = z / r_{2}
$$

After that, we apply DBSCAN~\cite{DBSCAN} to identify dense regions of the transformed space as track candidates. This method achieves the score of~$\approx 0.21$.

The third notebook explores another kind of transformation that maps from Cartesian space into possible track parameters space. Let's consider a simple 2D example with circular tracks. Such tracks can be parametrized in polar coordinate system as follow:

$$
r = 2r_{0}\cos(\phi - \theta)
$$

where:
\begin{itemize}
    \item $r$ and $\phi$ : are coordinates of a hit in the polar system;
    \item $r_{0}$ and $\theta$ : are coordinates of a centre of a circular track in the polar system.
\end{itemize}
 
So the transformation of Cartesian coordinates of a hit to polar coordinates is defined as:

$$
\phi = \arctan(\frac{y}{x})
$$
$$
r = \sqrt{x^{2} + y^{2}}
$$

\begin{figure}[ht]
\includegraphics[width=0.9\textwidth]{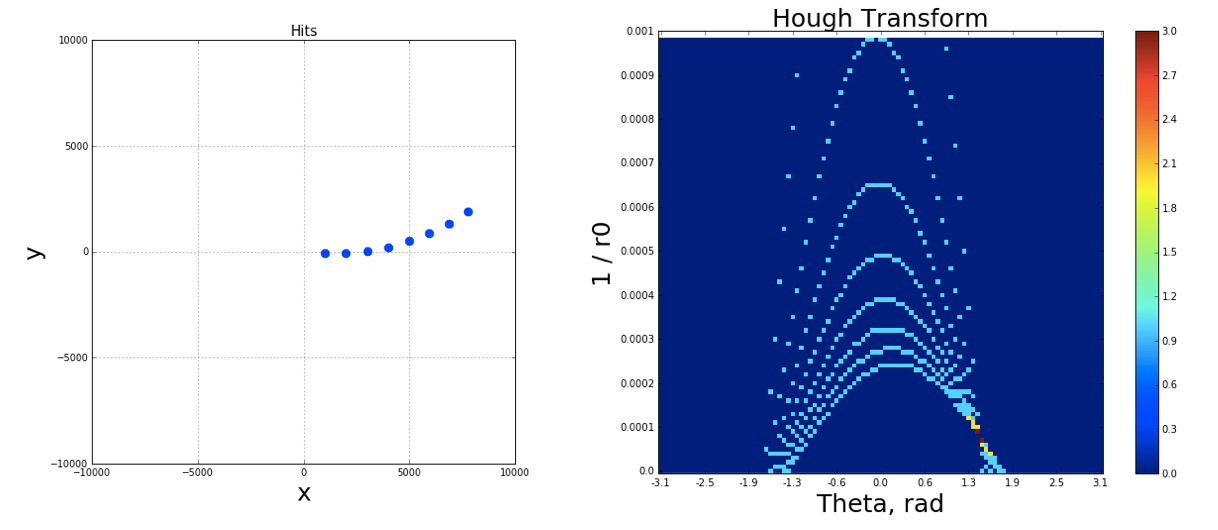}
\caption{Hough hit coordinates transformation: original (left) and transformed (right).}
\label{fig:sk_hough_transform}       
\end{figure}

In that system a linear track corresponds to the $r_{0} = \infty$. The Hough Transform~\cite{hough_transform} maps a hit in $(r, \phi)$ space to a curve in $(\frac{1}{r_{0}}, \theta)$ space (see light-blue curves on the right side of figure~\ref{fig:sk_hough_transform}):

$$
\frac{1}{r_{0}} = \frac{2\cos(\phi - \theta)}{r}
$$

A linear track in this space corresponds to the $(0, \theta)$ point.
For 3D case, we can parametrize a helix by cylindrical coordinates: $\phi$, r, z. For the simplicity, we can assume

$$
\gamma=\frac{z}{r}=const
$$

which is true for high-PT tracks. So we can map every hit into possible helix coordinate space as depicted by the figure~\ref{fig:sk_hough_transform}.

Afterwards, we split the track-parameter space into bins, and those with the higher density correspond to the higher possibility of tracks with the corresponding parameters and eventually represent recognized tracks. With the help of Hough transform one can achieve similar score as with DBSCAN ($\approx0.20$). 

Those approaches are intended to give an easy-to-grasp way of dealing with the data and help to develop geometrical intuition for the challenge.

\section{Competition Facts and Figure}
\label{s:competition}

\begin{figure}[ht]
\includegraphics[width=\textwidth]{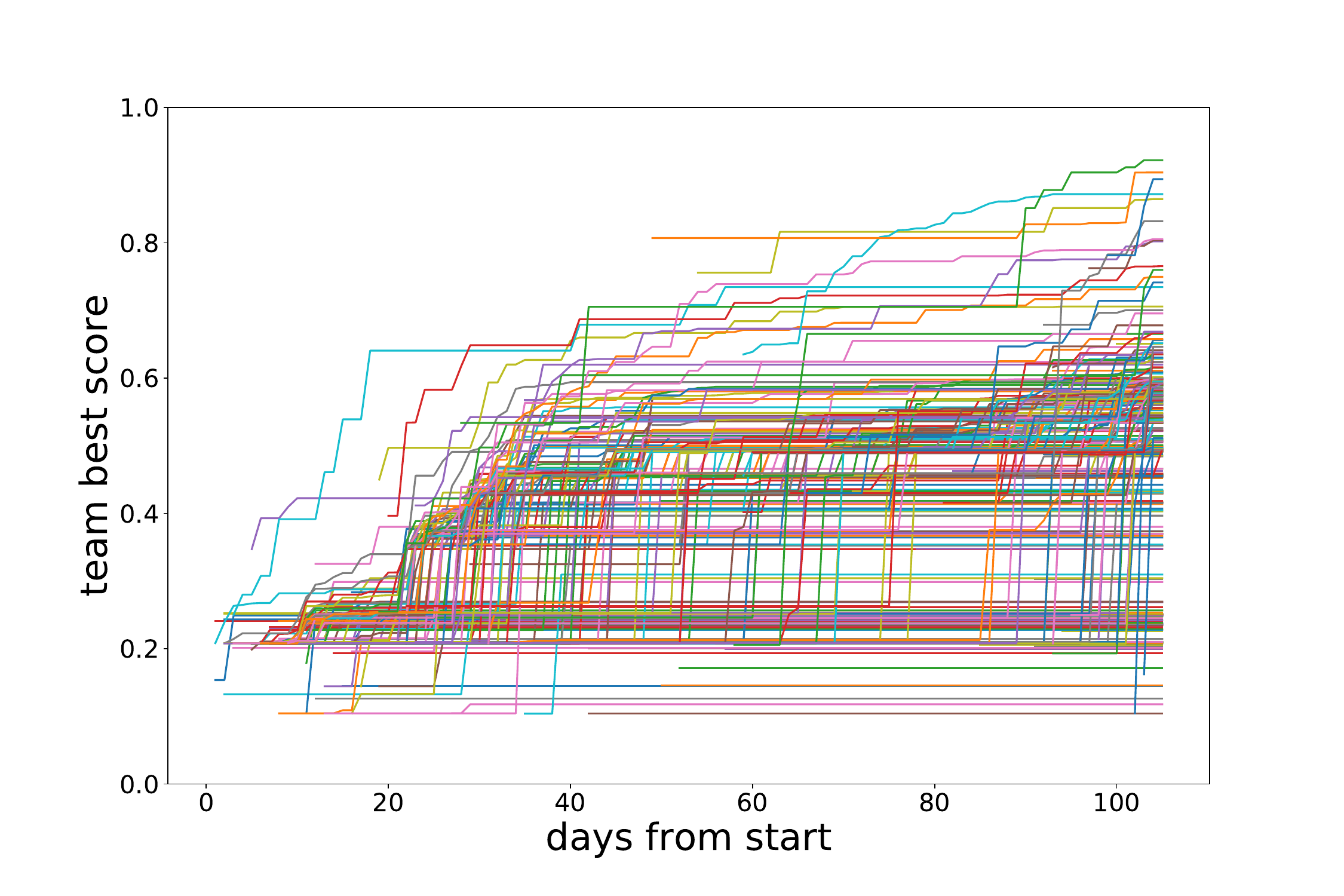}
\caption{Evolution of the best score of each team as a function of time.}
\label{fig:ScoreEvo}       
\end{figure}

The first phase of the TrackML challenge was well attended, with a total of 656 participants.
Fig.~\ref{fig:ScoreEvo} shows the evolution of the leader scores over the duration of the competition.
Several features are visible. 
First, a score of more than 90\% was only reached in the last days of the competition.
Also, there is a large cluster of candidates achieving a score of 20\% to 25\%, which corresponds to the 22\% performance of the DBSCAN starting kit described in section~\ref{s:startingkit}.
After around 30 days, public kernels achieving a performance greater than 50\% were posted on the public forum, which lead to a second group of candidates reaching a performance of 50\% to 60\% after 40 days of competition.
Finally, except for those groups, it is interesting that the candidates achieving the best performance are well separated from those groups and from each other.

\begin{figure}[htp]
\includegraphics[width=\textwidth]{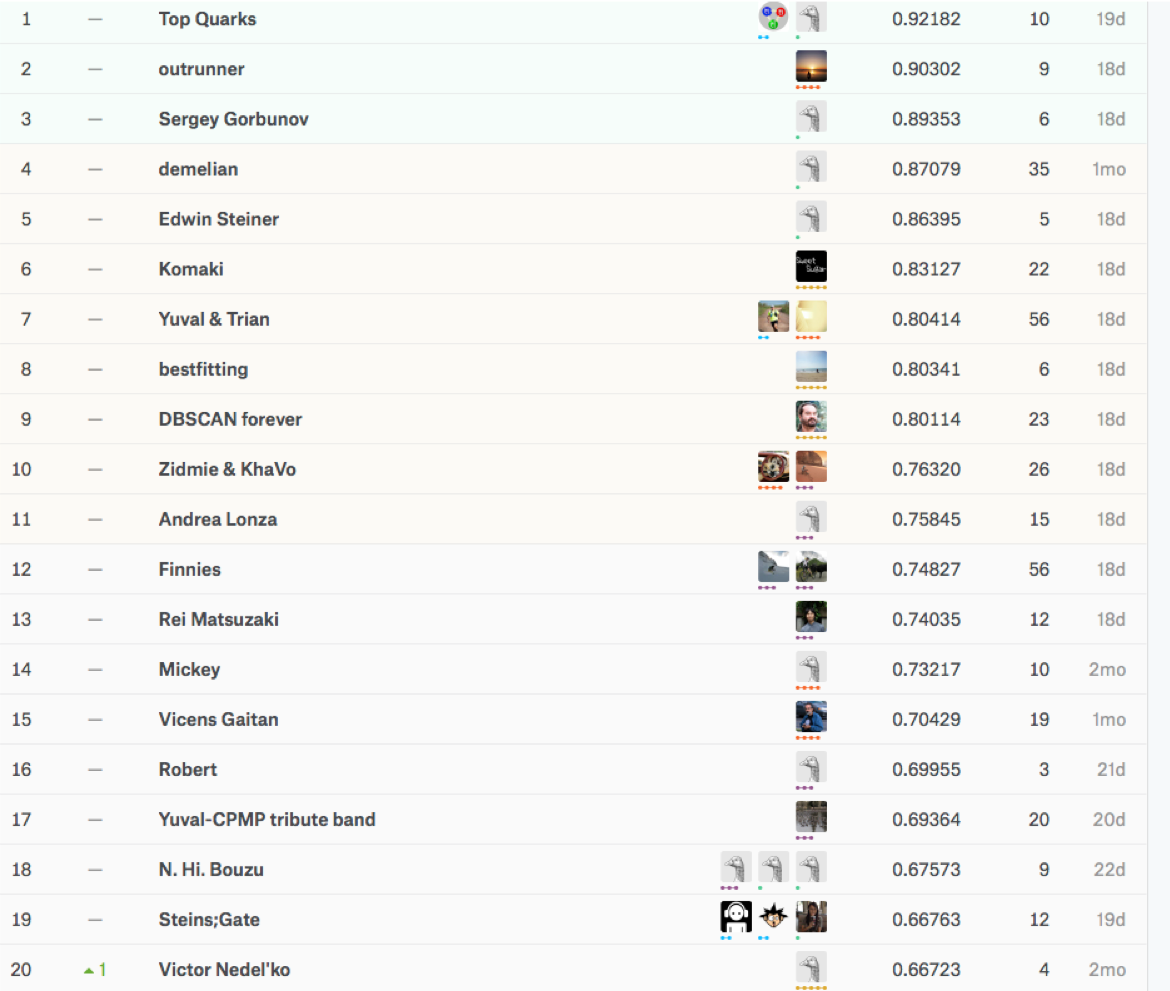}
\caption{Final leaderboard from the competition website.
\cite{TrackMLKaggle}
The various columns correspond to the final rank, ranking difference with respect to the private leaderboard (in this case identical except for the last candidate), candidate name, logo, final score on the 125 test events, number of submissions, and time elapsed since the last submission.
}
\label{fig:LB_Caps}       
\end{figure}

Fig.~\ref{fig:LB_Caps} shows the final leaderboard (LB) with the top 20 participants.
The dashes on the left column indicate that the final leaderboard ranks (determined on the 89 events of the private leaderboard dataset) are identical to those from the public leaderboard (determined on the 36 events of the public leaderboard dataset) for the ranks 1 to 19.

\begin{figure}[ht]
\includegraphics[width=1\textwidth]{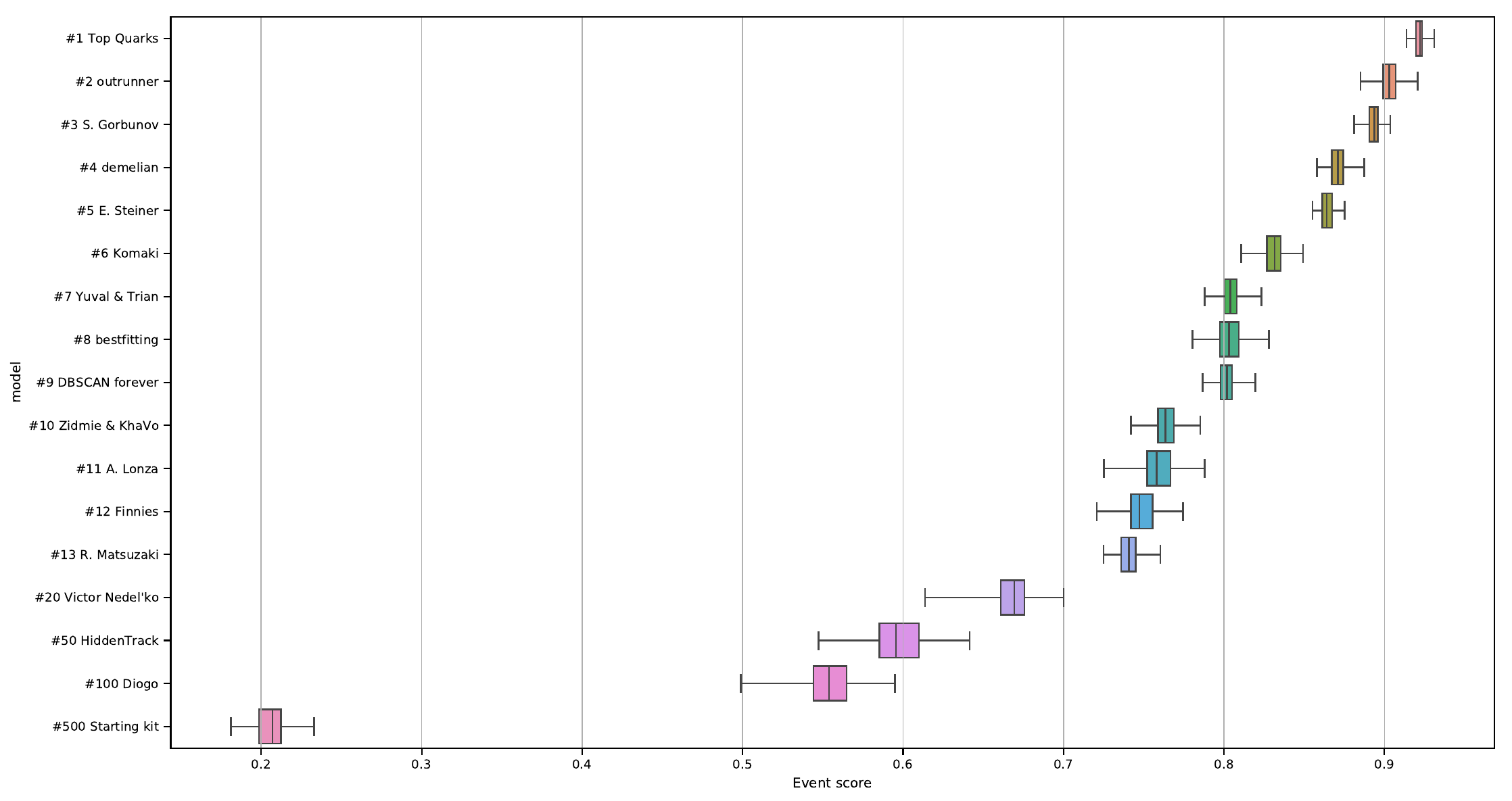}
\caption{Boxplot distribution of the best candidates and number 20, 50, 100, and starting kit. 
The box plots display the quartiles and extrema of the algorithm scores on the 125 test events.
A clear separation between the candidates is seen, validating the ranking.
}
\label{fig:Boxplot}       
\end{figure}

A deeper analysis of the score statistical accuracy has been done using the participants submission for the 125 test events.
Fig.~\ref{fig:Boxplot} shows the score distribution for the best participants;
While there is some overlap between the distributions, the means are well separated, except maybe for rank 7, 8 and 9.
A quantitative analysis has been done using the Wilcoxon signed-rank test \cite{wilcoxon}, a standard non-parametric statistical hypothesis test used to compare repeated measurements on a same sample to assess whether their population mean ranks differ.
The test is done on the 36 events of the smaller public dataset (which gave the rankings on the Kaggle web page), and gives a two-sided p-value; 
the only candidate pairs having more than  $1^{-6}$ are 10 and 11 ($3^{-6}$), 7 and 9 ($3^{-5}$), and more relevant 7 and 8 (0.03) and 8 and 9 (0.08).
All values are at the few percent level or (way) below, showing that the final ranking could not have changed due to statistical fluctuations.

\section{Study of algorithms performance}
\label{s:perf}

In this section, we analyze the quality of the top algorithms in more details beyond just the score, in particular with an eye on the criteria from a physicist point of view.

As a post-analysis of the competition, we have downloaded from the Kaggle website the submissions of the top candidates, and compared them to the ground truth. This allows us to know where each hit of each of the 125 test events, was correctly associated, and whether each track was correctly reconstructed (we consider a track to be correctly reconstructed if it contributes to the score as described in section~\ref{s:scoring}).
In addition to check the correct hit to track assignment, what is usually done in particle physics is to derive from the hits the particle parameters at the origin, and compare them to the ground truth;
this is an on-going study not reported here.

\subsection{Tracking efficiency}
\label{s:trackeff}

{\it Tracking efficiency} is commonly defined in particle physics as the {\it probability to reconstruct a track}. A good tracking algorithm must provide consistently high efficiency over a wide range of track parameters.

\begin{figure}[ht]
\includegraphics[width=1\textwidth]{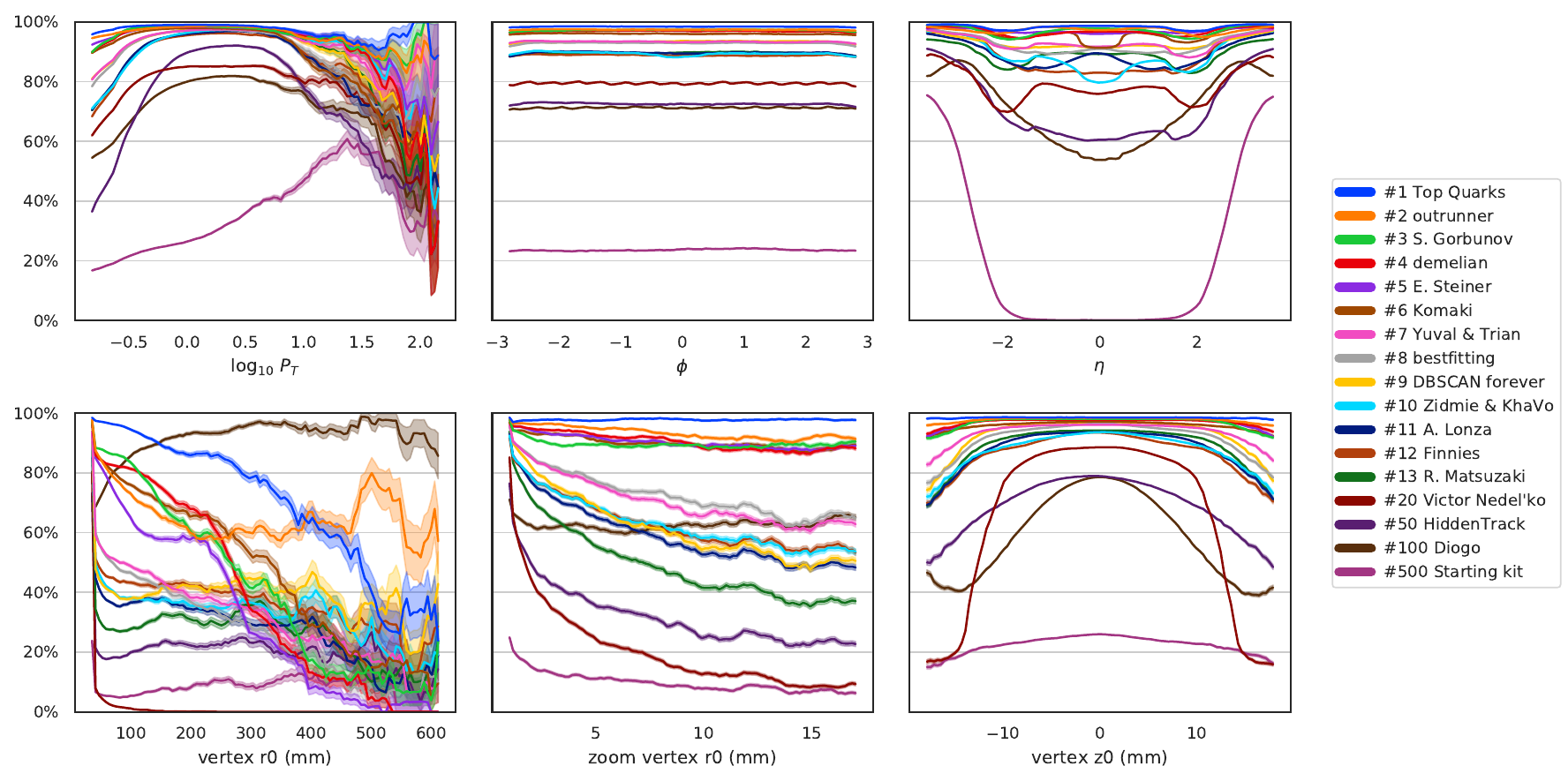}
\caption{Primary particle reconstruction efficiency for various candidates as a function of six physics variables.
See Fig.\ref{fig:unweightedpdf} for the variables definition, and the underlying probability distribution function (lower statistics means greater uncertainty).
}
\label{fig:AllEff}       
\end{figure}

The efficiency of a good algorithm is expected to be independent of the azimuth angle $\phi$ and the pseudo-rapidity $\eta$, and perform best for tracks originating from a low radius ($r_0$, that is, near the beam axis, where the primary particles are created.
Such distributions are shown for representative solutions to the TrackML challenge in Fig.~\ref{fig:AllEff}. 
The efficiency was computed using only the primary particles having created more than 4 hits (both expected result in well reconstructed tracks, and more relevant for physicists), hence the numbers can be slightly higher than those displayed on Fig.~\ref{fig:LB_Caps} and  \ref{fig:Boxplot}.
To put some perspective, to the 13 leaders have been added the contribution of number 20, 50, 100 and of the DBSCAN starting kit introduced in Section~\ref{s:startingkit} (500 is representative of the ranking it would have achieved in the competition).
The better the ranking, the higher and flatter is the efficiency distribution according to all variables, validating the choice of the scoring variable (see Section~\ref{s:scoring}).
One interesting exception is the candidate ranked 100, diogo, who achieves a very high efficiency (the best one) at large radius (see~Section~\ref{s:diogo}).
Those distributions have to be compared to the underlying probability density functions of the particles, which are indicated on Fig.~\ref{fig:unweightedpdf}.
It is necessary for an algorithm to be performing well where the bulk of the particles are.

An important algorithmic quality criterion is the ability to distinguish particles close to each other.
If a particle is created close to the axes origin, the $\phi$ and $\eta$ coordinates described in section \ref{s:detector} can be used to describe its direction of propagation.
The angular distance $\Delta R$ between two such particles separated by angles $\Delta \phi$ and $\Delta \eta$ can then be defined by:
$$ \Delta R = \sqrt{\Delta\phi^2 + \Delta\eta^2 } $$
For each particle, this distance is computed with respect to its nearest neighbour in the $\phi - \eta$ plane;
Particles of opposite charge sign (which bend in different directions) can be separated more efficiently, so the distance is also computed for the nearest neighbour of the same charge and of the opposite sign charge.
It is also relevant to ensure that the two particles were created close to each other, and a selection can be made on $\Delta z$, defined as the distance on the $z$ axis between the vertices that created both particles considered.

\begin{figure}[ht]
\includegraphics[width=\textwidth]{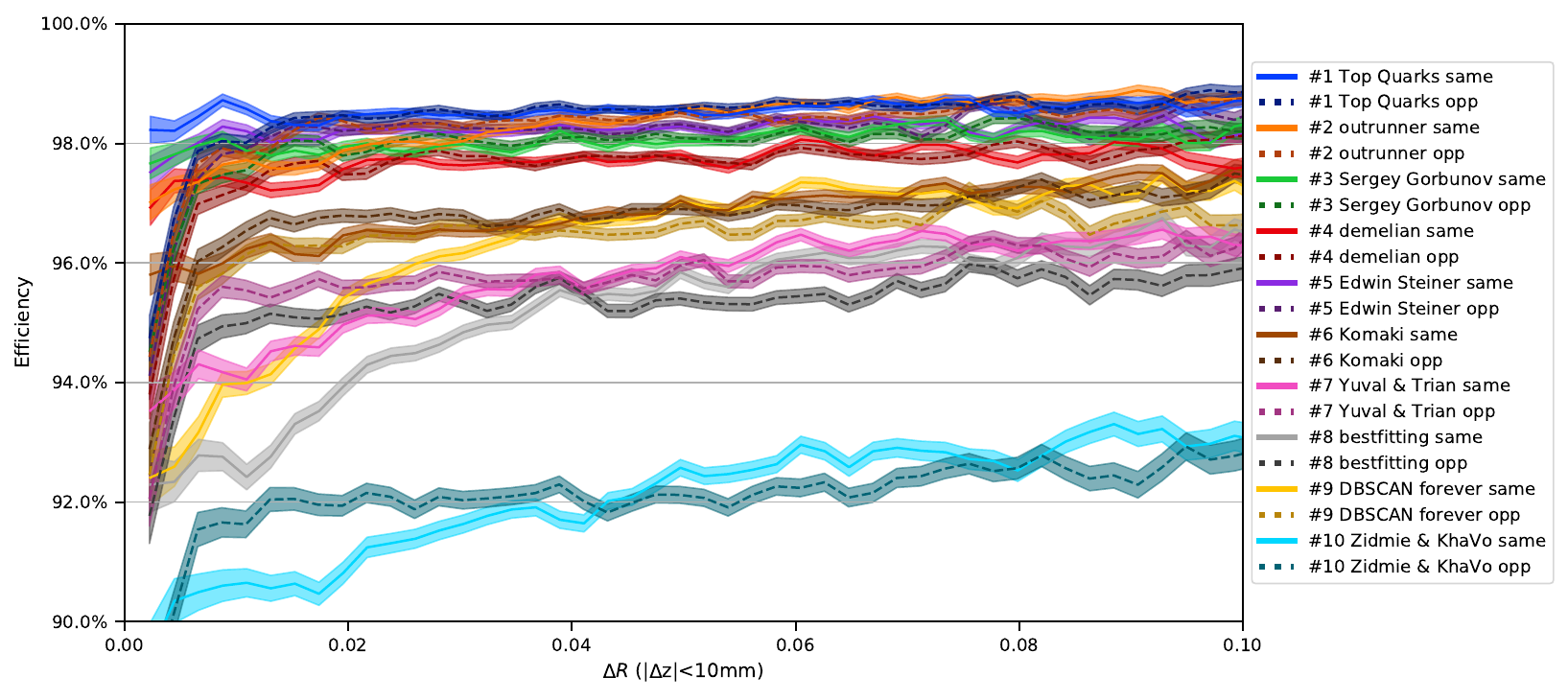}
\caption{Primary particle reconstruction efficiency for various submissions as a function of $\Delta R$, the nearest neighbour angular distance, with a selection $\Delta z < 10 mm$ on the nearest neighbour vertex z distance.
Each color corresponds to a different candidate;
full lines in bright tones correspond to nearest particle of same charge sign and dotted lines in dark tones to opposite charge sign.
}
\label{fig:deltaR}       
\end{figure}

Fig.~\ref{fig:deltaR} shows the efficiency as a function of $\Delta R$,  for the nearest neighbour having the same charge sign (full lines in bright tones) and the opposite (dotted lines in dark tones).
As expected, the efficiency drops when the angular separation is small, because of a more likely confusion between the points of the two particles. The behaviour differs according to the relative sign of the nearest neighbour. Minimizing the drop of efficiency at small $\Delta R$ is important because it would impact the detection of interesting phenomena yielding a bunch of almost collinear particles. This was not included specifically in the score, and we see that even the best algorithm, top-quark, suffers from this drop, albeit in a limited manner, as do the algorithms currently used in HEP. Different algorithmic choices reflect in different sensitivity to the closest particle.

\subsection{Track purity}
\label{s:trackpur}

In this section we explore the reasons why some tracks had a null score.
For example, did they fail to get a sufficient fraction of the underlying particle?
Did they aggregate several of them?

Following the definition of particle purity and track purity in section~\ref{s:scoring}, tracks can be categorized as follows:
\begin{itemize}
    \item {\bf Good} : Both purities above 50\% (the only case taken into account in the score)
    \item {\bf Split} : Particle purity below 50\%, track purity above 50\% (only a small part of the underlying particle was "caught")
    \item{\bf Multiple} : Particle purity above 50\%, but track purity below 50\%  (typically, the track aggregated the result of two different underlying particles; this is rare since most particles originate near the origin)
    \item {\bf Bad} : Both below 50\%
\end{itemize}

\begin{figure}[ht]
\includegraphics[width=\textwidth]{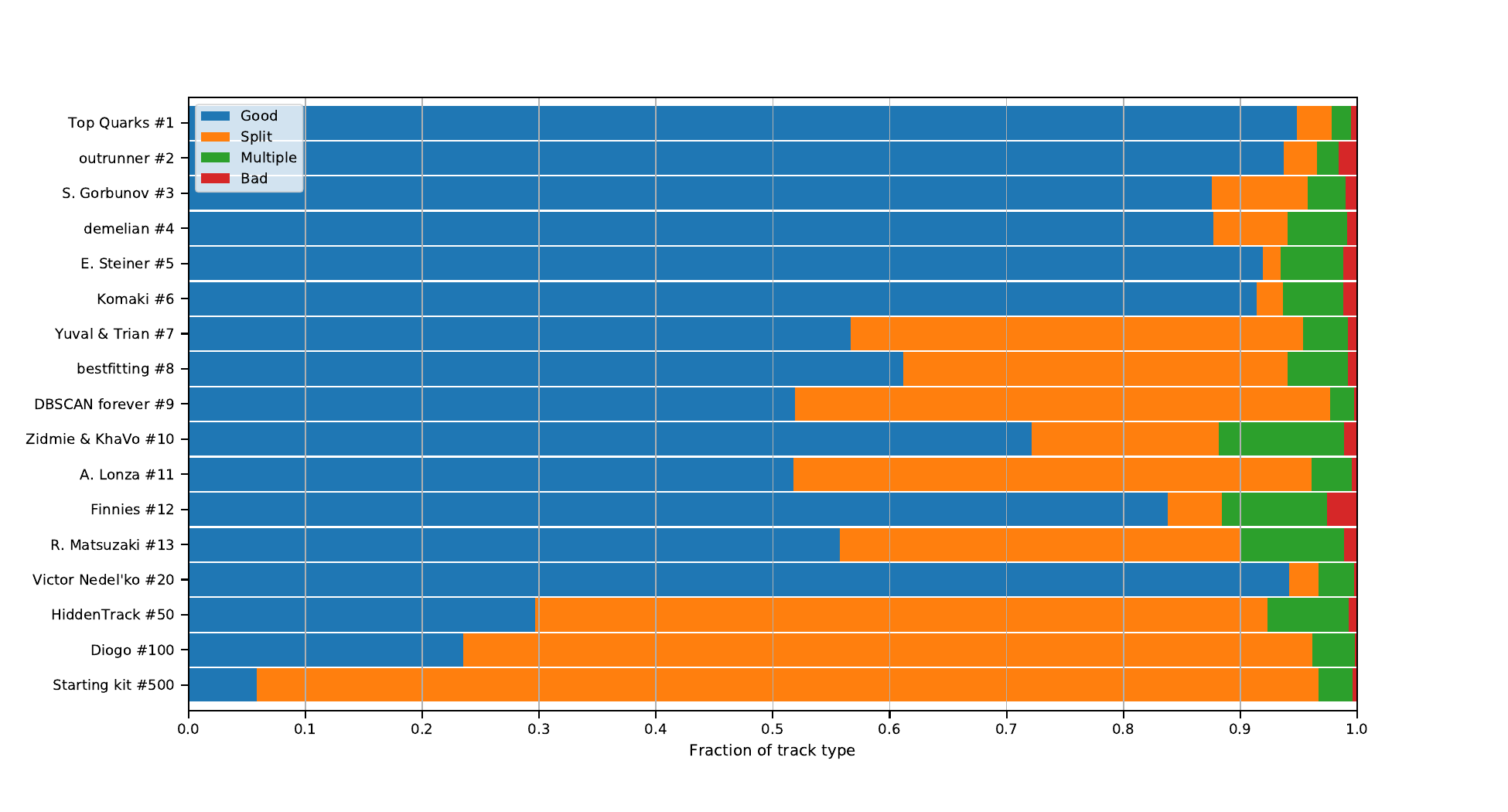}
\caption{Distribution of the kind of tracks for each of the algorithms under consideration.}
\label{fig:TrackType}       
\end{figure}

Fig.~\ref{fig:TrackType} shows the classification of tracks for the primary particles of the aforementioned algorithms.
Whereas the fraction of good tracks closely follows the global efficiency, it is interesting to note that for instance algorithm number 20 has an excellent fraction of good tracks among the ones selected, indicating that the "bad tracks" category must contain few tracks with a high number of garbage hits.

\begin{figure}[ht]
\includegraphics[width=\textwidth]{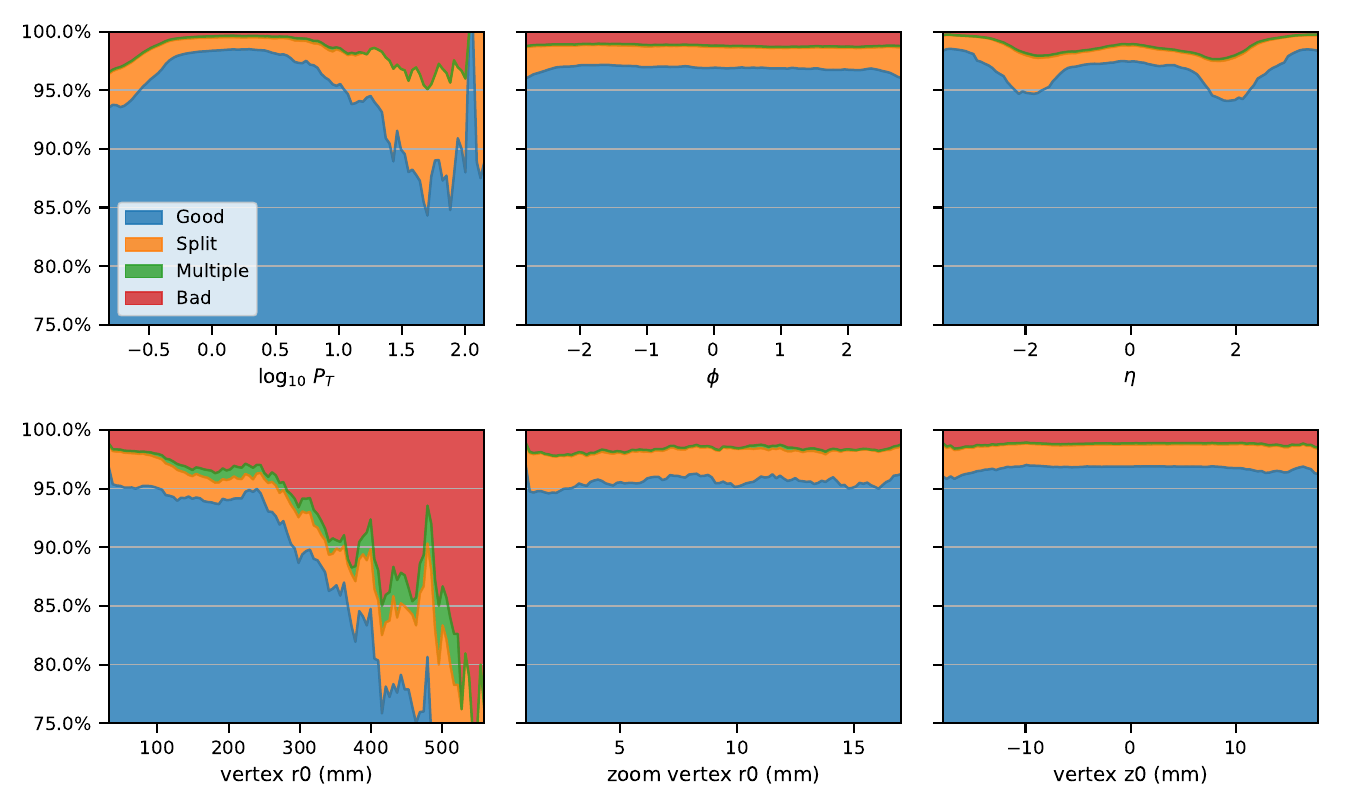}
\caption{Reconstructed track type distribution of the six physical variables of Fig.~\ref{fig:AllEff} for the TopQuarks contribution.
Note that the "good" category matches closely the efficiency of the candidate (although on Fig.~\ref{fig:AllEff} we only considered primary particles, slightly increasing the score.)}
\label{fig:TopQuarkTrackType}       
\end{figure}

The distributions of track types according to the six usual physical variables are shown in Fig.~\ref{fig:TopQuarkTrackType} which displays the track type distribution for the leader.
The "good" tracks represent the majority of the event, and the score reflects quantitatively the efficiency of Fig.~\ref{fig:AllEff} (which only took into account primary particles).
The amount of good tracks found is lower when going further from the origin, but is flat up to 300 mm in $r_0$ and along the $z$ axis coordinate; this was the secret behind the best rank.

\begin{figure}[ht]
\includegraphics[width=\textwidth]{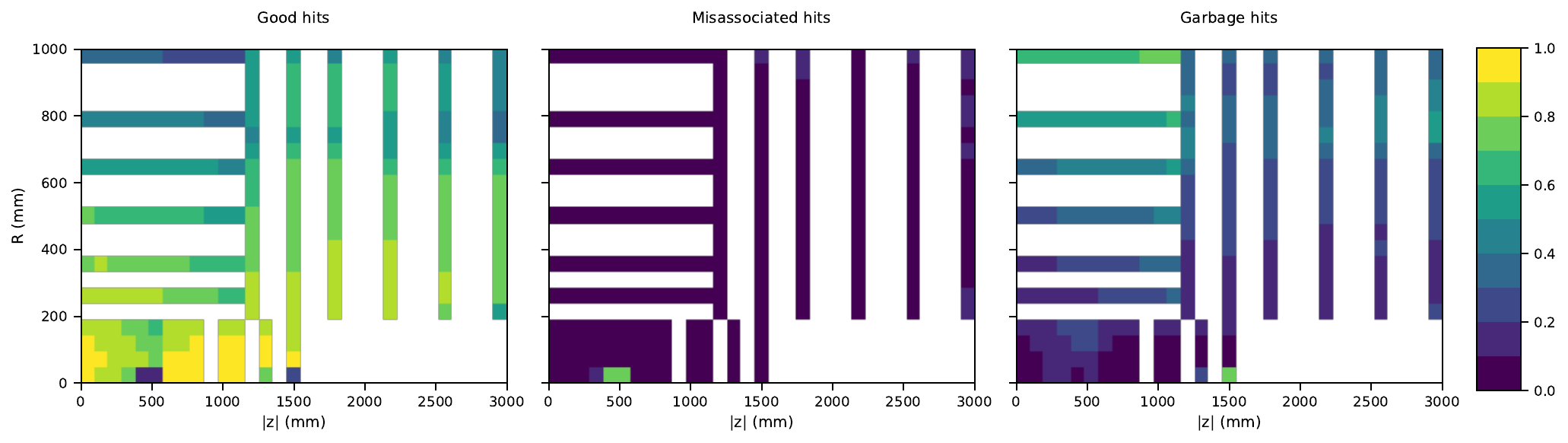}
\caption{Fraction of the good (left), mis-associated (center), and garbage (right) hits as a function or their position in the detector for the winning algorithm (TopQuarks).
}
\label{fig:HitMaps}       
\end{figure}

Finally, we use the hits to estimate where in the detector the algorithms perform the best.
We categorize the hits into three types:
\begin{itemize}
    \item If a hit was not associated to a good track, it is called \emph{garbage}.
    \item Otherwise, if it did not belong to the particle with the majority of the hits, it is called \emph{mis-associated}.
    \item Otherwise (good track and part of majority particle), it is called \emph{good}.
\end{itemize}
Fig.\ref{fig:HitMaps} summarizes the position of those types for the leading candidate.
We see that the algorithm performs best near the center, except for a bin of mis-associated hits close from the beam axis but far (around 50 cm) from the origin;
the garbage hits are mostly found on the detector outer layer (up to 80\% there).

\section{Selected algorithms}
\label{s:algos}
At the end the first phase of the challenge the participants were encouraged to provide a description of their algorithms to be able to claim their prize and enter the jury selection. 
As underlined in section \ref{s:perf}, one of the solutions has a rather atypical behavior and has an advantage for track with large impact parameter.
We first describe the algorithms underlining the overall strategies and innovations.
We then highlight interesting aspects that would deserve special attention in further work on tracking using machine learning.

\subsection{Algorithm Description}
All algorithms share common approaches and methods, but for the sake of clarity of the description, each algorithm is  described separately below leading to some unavoidable repetitions.

\subsubsection{Challenge Winner : top-quarks }
\label{s:topquarks}
{\it The team topquarks is composed of Johan Sokrates Wind ¬´ icecuber ¬ª and Erling Solberg "erlinsol". The technique was essentially developed by Johan Wind who is an  Industrial Mathematics master student in Norway.}

\begin{figure}[ht]
\centering
\includegraphics[width=0.45\textwidth]{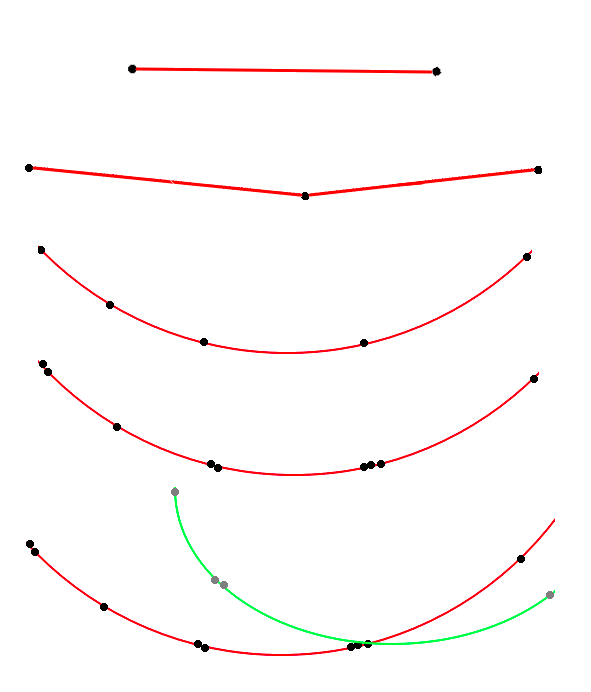}
\caption{Schematic representation of the steps of the challenge winner algorithm. From top to bottom : pair finding, extension to triplet, extension to tracks, addition of hits from overlapping modules, and final track disambiguation.}
\label{f:topquarks_steps}
\end{figure}

The winning algorithm\cite{github_top-quarks} is  modular so as to allow for efficient testing new ideas at all levels of the algorithm.
The algorithm's track following strategy is similar to that of several tracking algorithms currently used in production by high energy physics experiments.
The subsequent steps are instrumented with intermediate measurement of the quality of the solution. 
This allows to keep track of loss in score and was probably a key factor in winning the challenge.
While most of the code is written in C++/C++11, some training was performed using python for practicality.
A dedicated data structure was implemented especially for the challenge to favor many look-ups which could have been otherwise prohibitively expensive.
The compromise between hit combinatorics and search space usually present in standard track following algorithms as the algorithm proceeds with finding hits that belongs to track, is avoided in this method by fixing the branching in the combinatorial tree using an estimation of the density of polluting hits as a function of the hit location in the detector.
Binary classification is used at two stages of the algorithm and was trained on a very small number of events, as each event anyways contains a large number of hits and tracks.
The whole algorithm contains several parameters which turn out to be hard to optimize by trial and error.
The solution proceeds as follows, and as represented in fig \ref{f:topquarks_steps}
\begin{enumerate}
    \item {\bf Seed generation:} 50 pairs of layers are selected in the innermost part of the detector and all pairs of hits are created. A logistic regression classifier is trained on hit pair features, and allows to reduce the number of wrong seed, keeping a almost all good seeds.
    
    \item {\bf Extension to triplets:} using a straight line extrapolation from the pair of hits to the next layer allows to find compatible hits. The 10 closest hits are used to form triplets. A logistic regression classifier is trained on triplet features, and retains most good triplets, rejecting bad triplets.
    
    \item {\bf Track Following:} further hits are attached to track candidates, starting from the triplets, by running an helix extrapolation to the next layers using the last three hits of the track candidate. The closest hit to the crossing point if added to the candidate. This helix extrapolation is performed using a data driven estimation of the magnetic field so as to be more accurate.
    
    \item {\bf Track Consolidation:} the track candidate building so far has not taken into account overlapping modules which may lead to multiple hits per layer. Extra hits on each layer already crossed are added to the candidate if they are closer than a threshold. 
    
    \item {\bf Track Ambiguity Resolution:} the procedure so far as created candidates with potential overlap. Ambiguities are lifted by selecting the candidate with the least amount of estimated polluting hits (calibrated using training data), promoting it as final track and recursively removing its hits on all candidates.
\end{enumerate}

\subsubsection{Challenge Runner-up : outrunner }
\label{s:outrunner}
{\it Pei-Lien Chou "outrunner" is  a software engineer in image-based deep learning in Ta\"iwan.}

The solution that ranked second in the challenge is using an artificial neural network model predicting the adjacency matrix of hits in a all to all connection scheme.
A three hits helix compatibility check is used the post-processing 
The solution implement in python is organized as follows
\begin{enumerate}
    \item {\bf Adjacency Matrix prediction:} all pairs of hits are considered and 27 features are constructed from its quantities. A neural network model composed of  multiple wide dense layers is trained to predict the probability of the pair to be on the same track. There is a large class imbalance in the problem due of the predominance of pair of hits that are not belonging to the same track. This is overcome by sampling pairs from the negative class closer to the positive pairs to better define the boundary between the two classes. The accuracy weighted by the class cardinal is a better estimator of the performance of the model in this heavy imbalanced setup.
    \item {\bf Adjacency Navigation:} For a given initial hit, the pair with the highest predicted score defines a seed. The third hit that maximize the sum of probabilities of the pairs formed with the other two hits is considered for addition to the track. The compatibility of this third hits with an circle passing from the origin and the other two hits is used to definitely add or reject the hit to the track candidate. Once no hit remains as possible candidate for addition, and new initial hit is taken to create a new candidate.
    \item {\bf Track Merging and extension:} the quality of tracks is quantified by the amount of hits uniquely assigned to it. This quality is used to order overlapping tracks and assign hits to final tracks. Track candidates with high quality are extended by navigating the adjacency matrix with looser constraints.
\end{enumerate}

The proposed approach is an unstructured track following algorithm where the next hit is not provided by track extrapolation but directly a hit index based on the hit pair classifier score.
The reported prohibitive computation cost seems to indicates that much too many branches of the combinatorial tree are followed during the track following step.
Some level of tuning could result in similar performance without much loss of accuracy, and may constitute an alternative the the combinatorial track finder approach.
Although the proposed solution is prohibitively expensive to be used for making track candidate prediction, it is rather accurate, and further development might make it much more tractable.
It should be noted that graph based neural network approaches, such as the one presented in \cite{heptrkx} can combine the hit pair classification and the navigation of the adjacent hits.

\subsubsection{Challenge Second Runner-up : Sergey Gorbunov}
\label{s:gorbunov}
{\it Sergey Gorbunov is a physicist in Germany, expert in tracking software.}

The solution which ranked third  \cite{github_sgorbuno} 
during the challenge is following closely the strategy of most tracking algorithm in high energy physics.
A couple of novelties are introduced, the magnetic field is estimated from the data, instead of provided by a numerical model, the hit search on each layer is performed with a fixed grid lookout data structure, and tagging of candidate hits is done using incremental counts allowing for fast hit categorization.

\begin{figure}[ht]
\centering
\includegraphics[width=0.35\textwidth]{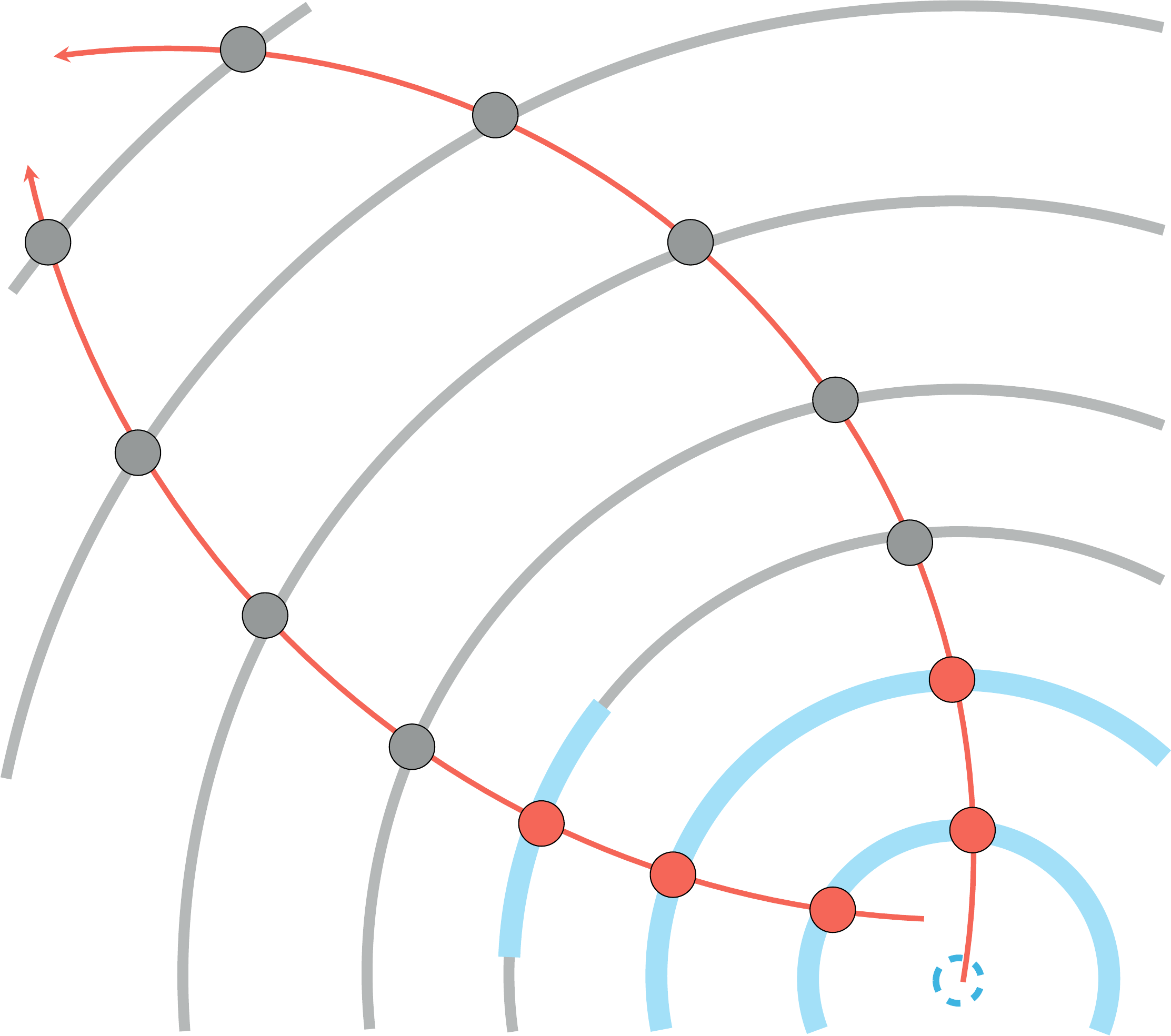}
\includegraphics[width=0.45\textwidth]{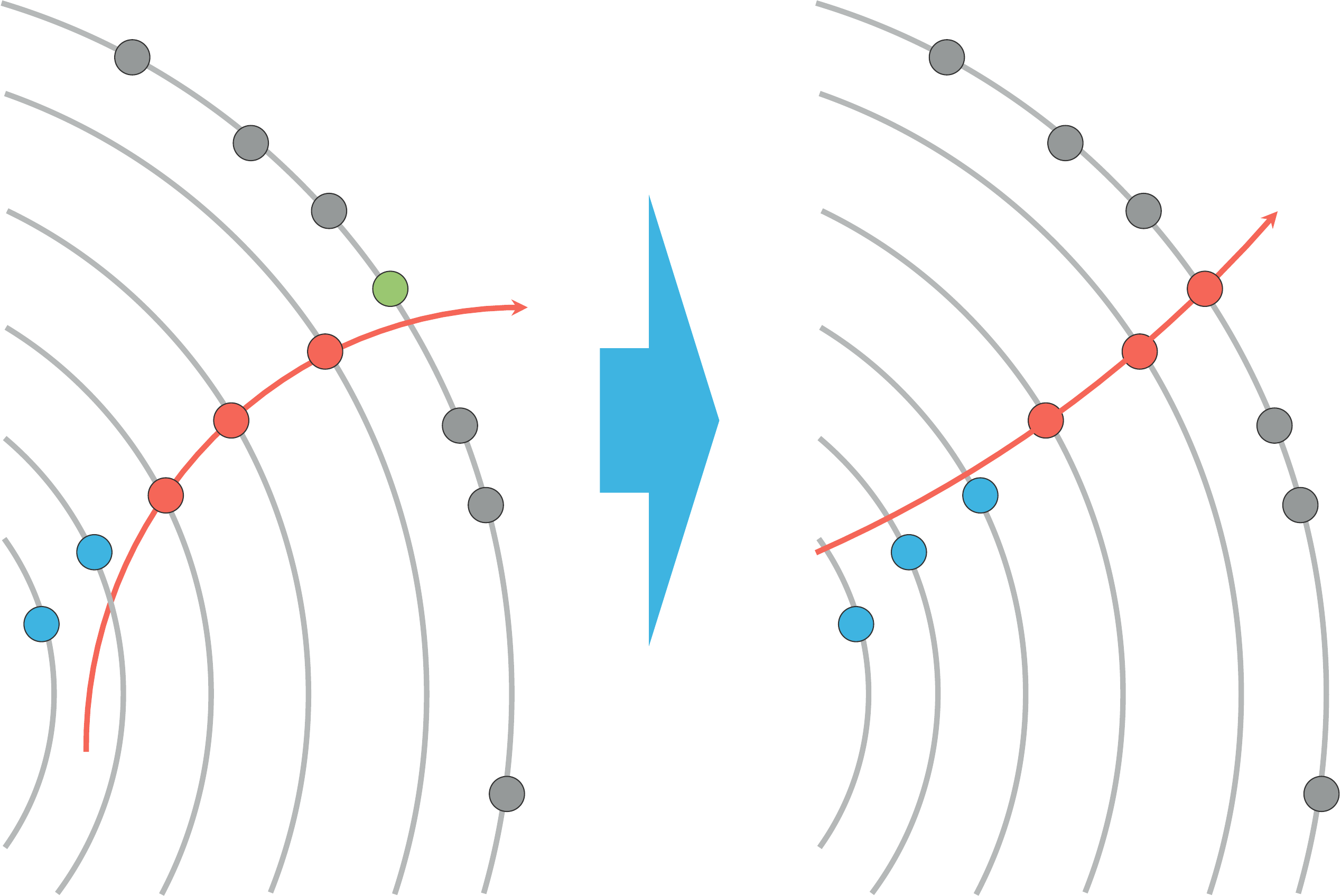}
\caption{Schematic representation of the steps of the challenge second runner up algorithm. Left : seeding from triplet of inner layers of the detector. Right : track candidate building using an helix extrapolation made from the last three hits of the candidate.}
\label{f:gorbunov_steps}
\end{figure}

The algorithm written in C++, proceeds as follow and as depicted in fig \ref{f:gorbunov_steps}
\begin{enumerate}
    \item {\bf Seeding:} triplet of layers of the detectors are determined to build up seeds from. All hits from the first layer are considered and a search for compatible hit is performed on the second layer, using a straight line extrapolation, from the origin and the first hit. Hits within a search window are taken as second hits. A second straight line extrapolation from the two hits onto the third layer is done to find hits within the search region. Triplets of hits are rejected based on their lack of alignment in the r-z plane.
    \item {\bf Track Candidate Building:} triplets found so far are taken individually and extrapolated through the successive layers of the detector. The last three hits of the candidate track is used for a local helix fit. This helix is used to find nearby hits on the layer of the last hit, to account for module overlap within layers. The helix is further used to propagate to the next layer, and the closest hit is added to the track candidate.
    \item {\bf Track Selection:} all the candidate tracks found from the original hit on the first layer are arbitrated to find a single best candidate. The candidate with the most hit of the least deviation from the local helix fits is retained and its hits are removed from the collection of hits. The algorithm restart with a hit on the first layer, and so on.
\end{enumerate}

The helix fit and extrapolations are performed using a data-driven estimation of an effective magnetic field, which takes into account variation in the magnetic field and the amount of material in the detector.
The good performance obtained with the absence of branching in the building of the candidate (one candidate per seed) is somehow surprising and an indication that the detector is granular enough compared to the amount of material such that there are little local ambiguities on hit association.

\subsubsection{Jury Innovative Prize : Yuval and Trian}
\label{s:yuvalTrian}
{\it Yuval Reina is an electronic engineer in Israel and Trian Xylouris is an entrepreneur in Germany }

The jury has selected the solution \cite{github_yuval_trian}  for the innovation in implementing a method that is not unlike the Hough transform \cite{hough_transform}.
The algorithm and the training of machine learning models are done in python.
The Hough transform in all 5 helix parameters at the origin is expensive computationally and prohibitive in dense collision environment like the one simulated for the challenge.
The smart indexing of the unique quintuplet of track parameters and the marginalization over transverse momentum and longitudinal impact parameter made the application of the Hough transform tractable in this case.
The algorithm contains the following steps
\begin{enumerate}
    \item {\bf Clustering:} assuming a doublet composed of the signed curvature and the longitudinal impact parameter of an helix, each hit correspond to a unique helix. This helix is parametrized by the polar angle, the sine and cosine of the azimuthal angle of the tangent on the longitudinal axis. The phase space of the triplet is discretized in order to form bins, bins in which hits of the same helix will accumulate. The helix triplet are smeared to prevent the fixed binning to limit the clustering, by allowing hit to migrate from nearby bins. Each bin is a track candidate if it contains at least one hit.
    The phase space of the initial doublets is discretized and scanned at random. Each set of track candidates are merged by assigning each hit, uniquely to the bin that has the largest cardinal.
    The indexing of each bin is made unique throughout the discretization of the five helix parameter, so as to effectively uniquely index a track parameter by the bin index.
    \item {\bf Ensembling:}
    Because the solution of the clustering of the previous step depends on the random walk in the doublet space, and the smearing on bin assignment, multiple passes of clustering are run and ensembled to produce a better solution.
    Each track index (or index of a bin in the five dimensions) determines a set of hits from which a set of track candidate features are extracted and then use to train a boosted decision tree for binary classification of the candidate.
    The multiple solutions of clustering are merged recursively in pairs, by selecting for each hit, the track with the best classification score.
    \item {\bf Post-Processing:}
    Further merging of candidates that are close in the helix parameter space is done by first estimating the curvature and longitudinal impact parameter that minimize the standard deviation of the remaining three helix parameters. The triplet of parameters are then recomputed and used to find hits with close-by helix parameters.
\end{enumerate}

This solution ranked seventh in the challenge, even though it does not include much domain knowledge about charged particle trajectories.
Machine learning is limited to a classifier used for selecting good track candidate from other clustering of unrelated hits.

\subsubsection{Jury Clustering Prize : CPMP }
\label{s:puget}
{\it Jean-Fran\c{c}ois Puget CPMP is a software engineer at IBM in France. He is a Kaggle competition grandmaster and a Kaggle discussion grandmaster.}

The jury has selected the following solution \cite{github_jfpuget} for bringing multiple improvements to the starting kit based on DBSCAN \cite{DBSCAN}. 
It uses a concept developed in a solution detailed in section \ref{s:diogo} for a definition of track quality.
Concepts of the Hough transform already discussed in \ref{s:yuvalTrian} are used to form track candidates in an iterative manner.

\begin{enumerate}
    \item {\bf Track parameter data bank:} the doublet : curvature and longitudinal impact parameters of all tracks observed in a training set are registered to be later used as possible track parameters to be scanned over. As such this can be seen as a track pattern data bank. 
    \item {\bf Track quality calibration:} the frequency of observation of quadruplets of crossed modules is measured from a training dataset, and is later used to estimated the likelihood of a track candidate.
    \item {\bf Track candidate building:} the space of track parameters doublet is scanned at random from the data bank of track pattern. For each track sub-parameter considered, all the hits are represented with three remaining parameters of the helix passing through the hit. A cluster of hits in this transformed space represent hits that closely share all five track parameters, and hence are considered a track candidate. This clustering is obtained using the DBSCAN algorithm in the hit transformed coordinate space. Ambiguities for hits already assigned to a track candidate are lifted by chosing to assigne to the track with the most hits and the best quality (based on frequency of module quadruplets).
    \item {\bf Ensembling:} the track candidate building above is ran once in the whole detector and once only in the most inner part of the detector. The two solutions are merged by virtue of track overlap.
\end{enumerate}

The use of DBSCAN, compared to the method detailed in section \ref{s:yuvalTrian} allows for a unbinned clustering in track parameters space.
By construction, this algorithm cannot find {\it loopers} (particles not escaping the tracker volume and returning to the beam line) and tracks from secondary vertex (tracks originating from a displaced pointt, hence missing the $z$ axis by a large distance).
The track quality estimator based on module quadruplet frequency is effective, but could be improved with a more granular and selective input.
In the transformation of the hit spacial coordinate into a triplet of track parameters, the observed non-uniformity of the magnetic field is taken into account by calibrating the transformation as a function of the longitudinal coordinate.

\subsubsection{Jury Deep Learning Prize : finnies}
\label{s:finnies}
{\it Nicole and Liam Finnie are software engineers in Germany. }

The jury has selected this solution\cite{github_jliamfinnie} for the use of recurrent artificial neural network (RNN), using long short term memory cells \cite{LSTM} (LSTM) also used in \cite{heptrkx}.
The DBSCAN algorithm reference in \ref{s:puget} is used to cluster hits in inner-most layers of the detector in order to produce tracklets seeds.
The recurrent network is used in place of a propagator to find the potential position of hits on subsequent layers of the detector.

\begin{figure}[ht]
\centering
\includegraphics[width=0.9\textwidth]{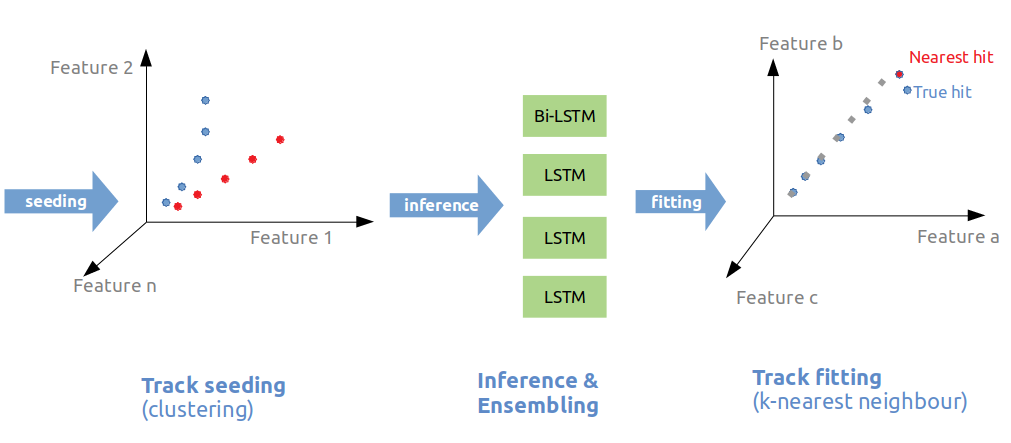}
\caption{Schematic representation of the steps of algorithm selected by the jury for its use of deep learning and recurrent network. Left : seeding is performed using DBSCAN. Middle : a recurrent model is trained and used to predict the positions of the next hits. Right : kNN-tree algorithm is used to find the closest matching hits.}
\label{f:finnies_steps}
\end{figure}

\begin{figure}[ht]
\centering
\includegraphics[width=0.9\textwidth]{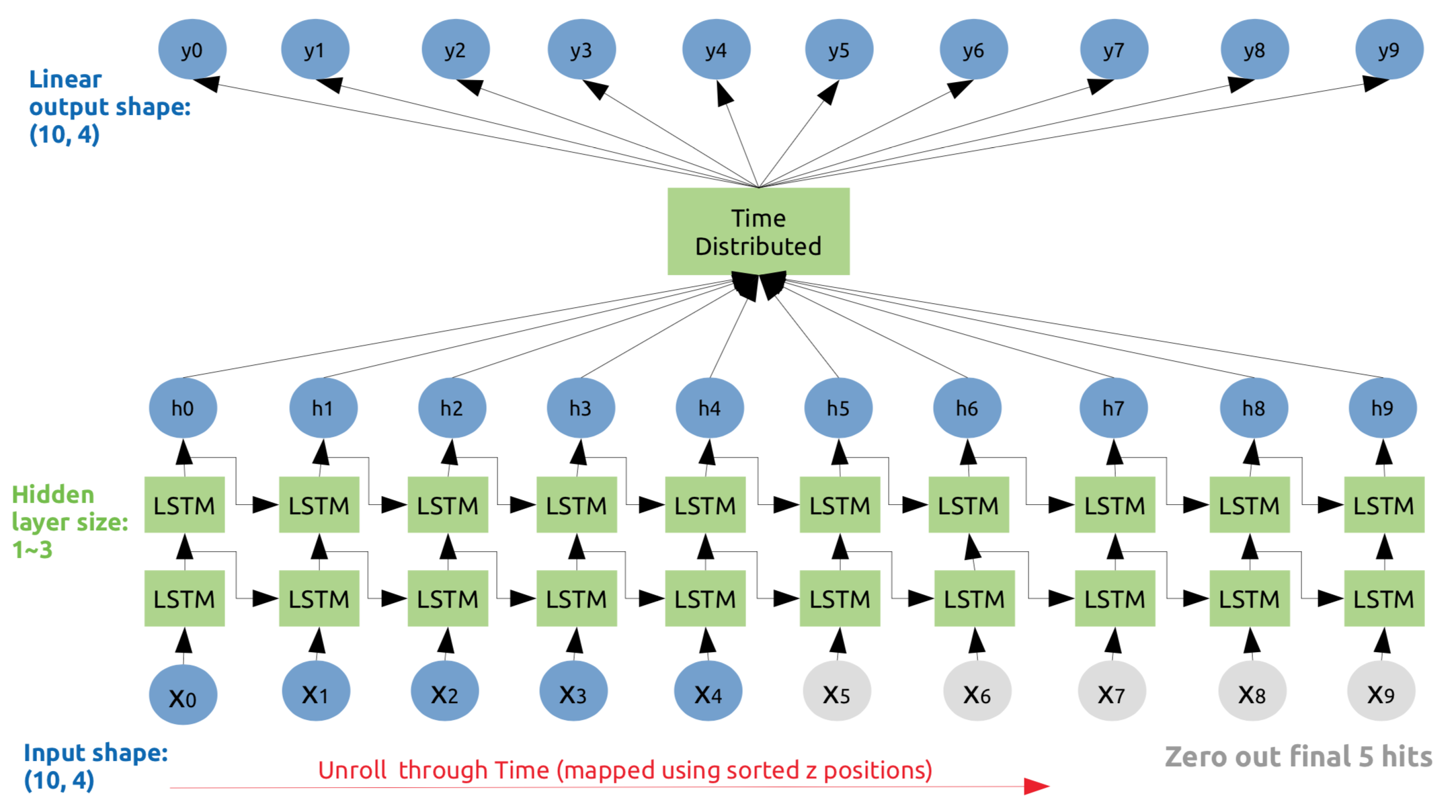}
\caption{Diagram of the recurrent neural network architecture used by the jury deep learning prize algorithm. A set of 5 hit quadruplets followed by 5 blank coordinates are presented in input to the model which produces in total 10 hit position quadruplets, that last five of which are used to look for matching hits in the detector. The model is a dual stacked LSTM with a dense model transforming the hidden representation into the space of hit position quadruplets. }
\label{f:finnies_LSTM}
\end{figure}

The team that ranked twelth in the challlenge also came up with the following algorithm implemented in python using keras \cite{keras} and Tensorflow \cite{tensorflow}.
It proceeds as follows and as depicted in fig \ref{f:finnies_steps}
\begin{enumerate}
    \item {\bf Seeding:} hits from all layers are considered in polar coordinates and clustered using the DBSCAN algorithm \cite{DBSCAN}. Each track candidate is truncated to the first 5 hits to produce a tracklet seed. The purity of the collection of seeds is improved using outliers rejection.
    \item {\bf Path Prediction:} from the observation that tracks are mostly straight lines in the coordinate systems ($\phi$, r) and (r, z), the ($\phi$, r, z, z/r) is chosen for track following. A recurrent unit is constructed (see figure \ref{f:finnies_LSTM}) with one hit position in input, and one hit position in output. It is ran along the 5 hits of the seed, and unrolled for 5 more iterations using zero-ed input to predict the position of the next 5 hits. 
    Multiple architectures for the recurrent model are implemented and trained separately. They are ensembled with averaging to provide the final prediction of the path of the charged particle in the detector.
    \item {\bf Hit association:} the k-D tree \cite{kdtree} is built using all hits of the events in the quadruplet space to efficiently find hits that are the closest to the path prediction, based on the Manhattan distance.
\end{enumerate}

Multiple architecture of the recurrent model are investigated, the training of the models is quite prohibitive to allow for a full optimization.
Computationally more economical recurrent cells such as gated recurrent units (GRU) could be used to make this training faster without a-priory loss of predictability.
This approach uses RNNs for track following and used the starting kit (see section \ref{s:startingkit}) to quickly get a set of good seeds. The algorithmic performance depends strongly on the seeding mechanism and could therefore be largely improved.
By design, this algorithm can only provide track candidates with ten hits. Variations of the model architecture and training could allow for shorter and longer tracks to be found.

\subsubsection{Organizer's Pick : diogo }
\label{s:diogo}
{\it Diogo~R.~Ferreira is a professor/researcher at the University of Lisbon, focusing on data science and nuclear fusion.}

As discussed in section \ref{s:perf}, one of the solutions drew the attention of the organizer as it performed quite uniquely well for track with large impact parameters (see bottom left plot of figure \ref{fig:AllEff}), regardless of the poor score overall (rank hundredth).
The solution~\cite{github_diogoff} uses a pattern matching algorithm also found in actual LHC trigger implementations that can be found in \cite{Sabes:2025864} and is based on the assumption that the training dataset contains all possible track pattern that can be observed in the detector during collisions.
The algorithm written in Python has the following two main steps.
\begin{enumerate}
    \item {\bf Route data-banks building:} from the observation that tracks are seldom sharing the ordered sequence of modules that are crossed, a set of routes are constructed from unique sequence of modules of the detector. A route consists of a sequence of module id and the expected position of the hit on the model. In the case of multiple tracks having the same sequence of module in the training data set, the prediction is made from the averaged positions. The weights of hits provided in the training data set are used in the averaging of hit position with a route, so as to favor the higher score.

    \item {\bf Hit matching:} routes that have at least one hit on each of its modules are used to build track candidates. In case of a candidate hit shared by multiple routes, the hit is assigned to the track candidate with the smallest average distance to the route predicted positions.
\end{enumerate}

This pattern matching algorithm performs poorly for tracks originating close to the beam line probably because of the initial assumption is incorrect for these type of tracks. 
This can be explained by the fact that a route is covering a non negligible finite volume in the space of possible tracks, and the density of track parameters along a route is too high and leads to ambiguities.
It however functions rather well for tracks created at secondary vertex (see section \ref{s:perf} for more details), likely because the density in the track parameter space is much lower, leading to unambiguous hit association within a nevertheless ambiguous route.

\begin{figure}[ht]
\centering
\includegraphics[width=0.8\textwidth]{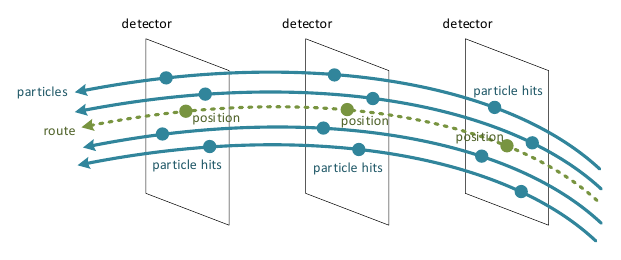}
\caption{Schematic representation of the route used in the algorithm of the "Organizer's pick". The route is precomputed from the positions of hits from tracks sharing the same sequence of traversed modules of the detector. Track candidates are produced by selecting the list of hits that are the closest to the predefined route. }
\label{f:diogo_route}
\end{figure}

\subsection{Lessons Learned}
\label{s:lessons}
It is not too surprising to find among the top ranking and winning solutions, algorithm highly inspired from the {\it seeding, track following, track selection} three-stages approach implemented in the current charged particle tracking algorithms.
The variations on this baseline approach are however interesting for future development of tracking algorithms.

\subsubsection{Accuracy Driven Steps}

The monitoring of the loss of accuracy at each step of the winning solution is probably a vector of its success in the competition as development an improvement are mainly guided by retaining almost all of the good track and hits.
Even though the metric of the challenge cannot be applied as such at each step and iteration, the participants were able to find useful and realistic proxies that help them develop their algorithms without losing the maximum score in objective.

\subsubsection{Data Driven Estimation of the Magnetic Field}

The model of magnetic field used in the simulation of the data for the challenge was not been provided in the data set description, as it was not deemed necessary and in order to simplify the data set publication.
Participants however observed loss of accuracy in their prediction, due to non-uniformity of the magnetic field.
Corrections which were applied have been derived in a data-driven manner, and would not only contain actual modification of the magnetic field, but also the amount of material that composes the detector.

Due to the imperfect modeling of the geometry of the real detector, the models for the magnetic field and material used in reconstruction software are only approximate and may lead to inaccuracies.
Such data-driven measurement of the magnetic field and geometry models may lead to better algorithms in the future.

\subsubsection{Computational Cost of Deep Learning}

The teams which applied deep learning to the vast amount of training data provided in the challenge had to face computation resource limitations.
Even with the use of general purpose Graphical Processing Units (GPUs) it took multiple days to train models.
As in many occasions, the phase space of hyper-parameters of such models were not fully scanned for the optimal set.
The participants did not report whether their proposed solution would perform better if given more resource and time.

On the other hand, however long it may take to train models and optimize the set of hyper-parameters, the use of the trained model can be extremely fast and potentially faster than conventional approaches.

\subsubsection{Hyper-parameter Tuning}

Deep learning methods are not the only algorithms, which have hyper-parameters that require tuning.
Many solutions proposed in the challenge were tuned {\it by hand} by the participants using their algorithm knowledge and intuition. 
A more systematic approach would probably require more computational resources, so as to fully the performance of the algorithm.
Further methods of tuning could involve Bayesian optimization using Gaussian processes regression of the performance function, or evolutionary algorithms which can be used to find an optimum functioning point of the proposed methods.

In the case of the challenge with a single score as a measure of goodness, such hyper-optimizations are rather easy to implement (at the cost of more resource).
On the other hand, in the context of charged particle reconstruction in an experiment software, there might not be a unique figure of merit of tracking.
Multiple quantities which have counter-balancing importance for the scientific throughput play a role, and the scientific throughput is not directly quantifiable.

\subsubsection{Noise-Driven Control of the Combinatorial Explosion}

The solution detailed in section \ref{s:topquarks} is inspired from the the canonical charged particle tracking algorithms.
In particular, during the track following steps, the number of candidate hits to be considered at each step is not controlled using the estimated error on the predicted position (which can be numerically expensive to compute), but by using the density of track outliers.

This approach may allow for faster software with an exact control on the complexity of the algorithm.
It might help to recover efficiency in pattern recognition of tracks which experienced large statistical fluctuations at some point of their trajectories through the detector.
The increased, controlled size of the tree would have to be balanced with a significant gain in computation to be beneficial of course.

\section{Conclusion and Outlook}
\label{s:conclusion}

The Accuracy phase of the Tracking ML challenge has introduced a variety of approaches, some of them being completely original for the field, as listed in section~\ref{s:lessons}. The quality of the algorithms is excellent, as it has been studied in depth in section~\ref{s:perf}. In particular, the fact that 99\% efficiency is reached over a wide range of parameters (see ~Fig.\ref{fig:AllEff}) indicates that the quality is similar to the  state of the art, although a direct comparison is not available at this stage.  This was not given for granted as using a single score as a factor-of-merit is very unusual in the domain. The challenge being a competition, and despite a very active discussion forum on ~\cite{TrackMLKaggle}, there was not a lot of collaboration between top participants. Thanks to the software released by the participants, more in-depth studies developing algorithms combining the different ideas have been launched.
    
From the domain point of view, the goal was to obtain new algorithms which are both of good quality and fast. For the first Accuracy phase, there was deliberately no incentive on the execution time, beyond what was practical for the participants. In the post competition survey, time between 10 minutes per event and one day per event have been reported. The Throughput phase has been launched on Codalab~\cite{TrackMLCodalab} in October~2018 till March 2019 and was still running at the time of writing. However, it appears already that some participants are managing to obtain very high score in just a few seconds, as a first hint of the success of the two stage approach. In depth analysis of the Throughput phase will be published in a future paper \cite{trackml_throughput}.

\section*{Acknowledgements}
The team would like to thank CERN for allowing the use of the dataset, and Kaggle for hosting it. We are very grateful to our generous sponsors without which the challenges would not have been possible. Platinum sponsors: Kaggle, Nvidia and Universit\'e de Gen\`eve. Gold sponsors: Chalearn, ERC mPP and DataIA. Silver sponsors : CERN Openlab, Paris-Saclay CDS, INRIA, ERC RECEPT, Common Ground, Universit\'e Paris Sud, INQNET, Fermilab and pyTorch.  TG acknowledges the support of the Swiss National Science Foundation under the grant 200020$\_$181984. SG acknowledges the support of the German BMBF ministry. This project has received funding from the European Union Horizon 2020 research and innovation programme under grant agreement No 724777 ``RECEPT'', No 772369 ``mPP'' and No 654168 ``AIDA-2020''. In addition, the organizers would like to thank participant Pei-Lien Chou "outrunner" for major contributions, Maggie Demkin and Walter Reade at Kaggle and  the members of the International Advisory Committee : Markus Elsing (CERN), Frank Gaede (DESY), Alison Lowndes (Nvidia), Maurizio Pierini (CERN), Danilo Rezende (Google DeepMind), Marc Schoenauer (INRIA-Saclay) and Svyatoslav Voloshynovskyy (U Gen\`eve).  

\bibliographystyle{spmpsci} 
\bibliography{trackml}

\end{document}